# Study on ploughing phenomena in tool flank face – workpiece interface including tool wear effect during ball-end milling

S. Wojciechowski [a], J. Krajewska-Śpiewak[b] R. W. Maruda [c], G. M. Krolczyk [d*], P. Niesłony [d], M. Wieczorowski[a], J. Gawlik[b],

*Corresponding author
[a] Poznan University of Technology, Faculty of Mechanical Engineering, Piotrowo 3, Poznan 60-965, Poland, E-mail address: sjwojciechowski@o2.pl
[b] Cracow University of Technology, Jana Pawła II 37, 31-864 Kraków
[c] University of Zielona Gora, 4 Prof. Z. Szafrana street, 65-516 Zielona Gora, Poland
[d] Opole University of Technology, 76 Proszkowska St., Opole 45-758, Poland

**Abstract**

Ploughing phenomena occurring during precise machining processes affect the formation of surface finish and progress of tool wear. Therefore, this study presents an evaluation of ploughing phenomenon by studying the ploughing forces in tool flank face – workpiece interface during precise ball-end milling of AISI L6 alloy steel. Developed original model of ploughing forces involved the effect of minimum uncut chip thickness $h_{min}$ and ploughing volume during ball-end milling. Estimated ploughing forces are sensitive to variations of surface inclinations and progressing tool wear. Moreover, intense growth of ploughing force during milling with worn tool is affected by irregular and non-circular profile of worn cutting edge below the stagnant point, and by the appearance of attrition and micro-grooves on tool flank face.

Keywords: ball-end milling; ploughing; tool wear; minimum uncut chip thickness,

1. INTRODUCTION

The circular profile of cutting edges enables application of ball-end mills in the precise machining of parts with a curvilinear surfaces. Therefore, the ball-end milling process is nowadays a very popular machining technology employed in many areas, including the aerospace industry and the production of wing parts or jet engine blades, as well as in the automotive industry to the production of panel molds, or the elements for a transmission systems [1]. The high restrictions on the quality of the curvilinear surfaces after the ball-end milling impose strict requirements concerning the selection of process input parameters. An important factor influencing the surface finish constituted during precise or micro ball-end milling processes is the ploughing phenomenon [2]. According to Tsybenko et al. [3], ploughing is a component of abrasion phenomenon, which in turn is the one of fundamental wear modes of ductile materials. Thus, the operational properties and strength of a machined surfaces can be influentially affected by the ploughing of a workpiece during the chip decohesion process. In addition, according to Zhou et al. [4] the ploughing occurring in cutting can affect the tool wear. The ploughing is manifested as the plastic deformation of material parts which are not removed from a workpiece during the cutting tool engagement [5]. During machining processes, the ploughing phenomenon is strictly characterized by a ploughing force, which appears during the cutting process with instantaneous uncut chip thicknesses contained in a range between the zero and $h_{min}$ [6]. In a recent years, the determination of ploughing during cutting has been the subject of study by some researchers. Wan et al. [7] determined the coefficients of a ploughing force in function of the volume of material extruded under the tool flank face. In order to determine the ploughing force coefficient, authors decomposed the static force measured during milling of aluminum alloy Al7050-T7451 into the ploughing and shearing force components. Chen et al. [8] evaluated the ploughing force during micromilling of KDP crystal. In order to resolve the total force to ploughing components and shearing components, the linear fitting of forces located at the rounded cutting edge was conducted to obtain the forces at the zeroth uncut chip thickness. Based on the conducted study, authors found that ploughing force increases rapidly with the increase of cutting edge radius. Moreover, the shearing force values are larger than values of a ploughing force. Popov and Dugin [9] studied the ploughing forces during chipping process of aluminum and alloy steels. Authors have compared the method which considers the extrapolation of forces to the zero uncut chip thickness with the experimental one in which the forces are being measured for a different tool wear values. Authors have found that values of ploughing forces estimated on the basis of the extrapolation method were substantially higher than values of ploughing forces obtained with the application of comparison method of total forces determined for a distinct flank wear values. Laakso et al. [10] studied the effect of ploughing on the feed force during longitudinal turning of AISI 304 stainless steel. Authors have applied the 2D FEM model implemented in DEFORM software and compared different friction models. On the basis of carried out research, it was found that measured values of ploughing forces are underestimated in relation to the simulated ones, which can be caused by a discrepancies in workpiece constitutive model and omitted residual strain from a subsequent cutting tool passes. Salehi et al. [11] estimated the cutting and ploughing forces during orthogonal turning of 1020 steel using an extended Kienzle model and Bayesian-Markov-Chain-Monte-Carlo (MCMC) model. It was demonstrated that application of Bayesian inference method can be employed to a successful prediction of ploughing and cutting forces with a minimized input data. Uysal and Altan [12] studied the ploughing forces during micromilling of CuZn30 brass. They applied the orthogonal cutting slip line approach considering a rounded cutting edge. Based on the research, it was found that effect of ploughing on thrust forces was higher than on the cutting forces and increased with the growth of cutting speed, rake angle and cutting edge radius. Bhokse et al. [13] estimated the ploughing forces during turning process of a hardened AISI 52100 alloy steel. Authors have calculated the ploughing forces with the application of Oxley's predictive machining theory. They observed that ploughing force has a 30% to 40% contribution in the whole generated cutting force during turning of a hardened steel. Nevertheless, within the lower cutting depths, the ploughing forces have higher values than shearing forces. Liu et al. [14] analyzed the ploughing force and $h_{min}$ during nano-machining of FCC single-crystalline material. Authors have conducted the molecular dynamics (MD) modeling and the nanoscratching tests to validate the analytical model with two distinct crystal orientations. Experiments have revealed that ploughing width can be estimated during nanomachining of FCC crystal, considering the uncut chip thickness, radius of cutting edge, cutting direction and surface crystal orientation. Luo et al. [15] proposed a 3D geometrical approach for a ball-end milling process to reliably estimate the ploughing volume of the chip. As part of this approach, authors have divided the theoretical chip into the large number of discrete elements in the range of ploughing and shearing regimes. It was observed that micro ball-end milling with relatively high feed rates contributed to the decrease of ploughing phenomena and ploughing volumes. The proposed approach can be applied for a prediction of ploughing volumes during micro ball-end milling with a complicated tool paths. Mishra et al. [16, 17] studied the ploughing behavior of the ellipsoidal asperity during sliding through the rigid-plastic workpiece. Authors have found that asperity size and shape affect the distribution of ploughing forces. This observation indicates that geometry of a cutting edge during machining process can have influential effect on the ploughing phenomenon.

The evaluation of ploughing requires the reliable estimation of $h_{min}$, since the ploughing forces are found during cutting processes with the instantaneous uncut chip thickness values lower than $h_{min}$. In recent years, the problem of minimum uncut chip thickness determination has been the subject of many studies. Liu et al. [18] proposed a criterion based on the Kragelskii-Drujuanov equation for scratch testing to predict the transition from ploughing to chip formation and incorporated it into the slip-line field model. Another extensively studied analytical approach is based on the definition of the stagnant point's location on the rounded cutting edge of the tool. At the stagnation point, the flow of ongoing material is divided into two directions: above its position material slides up the tool's rake face to form a chip; while material below its position is pushed down and forms the machined surface. Storch and Zawada-Tomkiewicz [19] used the principle of a stagnation point to determine the $h_{min}$ for orthogonal turning of C55 steel through analysis of unitary force increments on the cutting edge. Fang et al. [20] proposed a taper cutting experiment on monocrystalline silicon where uncut chip thickness increased along the cutting path. Through topographic analysis of the machined groove, it was noted that material was not being removed immediately after the tool-workpiece contact was initiated, but after the instantaneous uncut chip thickness reached a transitional value, taken as the $h_{min}$ parameter. Gawlik et al. [21] and Shi et al. [22] adapted analysis of acoustic emission (AE) signals in micro-milling experiments to estimate the onset of cutting. Researchers have noticed that the magnitude of the root mean square AE signal in the ploughing regime is very low (almost zero). In case when the instantaneous uncut chip thickness value reaches the $h_{min}$, the AE signal picks up due to noise generated by crack propagation and increases in proportion to the chip thickness. Rezaei et al. [23] carried out comprehensive experimental trials to evaluate how cutting parameters and lubricating systems affect chip formation in micromilling of titanium alloy. In their work, for every set of cutting parameters, chip morphology, average surface roughness parameter $Ra$ and feed marks on machined surface were compared. They concluded that empirically $h_{min}$ is not a singular value but a range of values. In the work of Oliviera et al. [24] a similar approach was used for micromilling of AISI 1045 steel which involved simultaneous analysis of chip morphology, surface roughness parameters and feed mark topography to estimate the $h_{min}$ range. With the aid of numerical simulations, the $h_{min}$ value could be determined through analysis of chip separation criteria from the effective rake face by employing a series of simulations with discretely selected uncut chip thickness to radius of cutting edge ratios ($h/r_n$). When simulation results are evaluated in increasing order of $h/r_n$ parameter a $h_{min}$ is established for the uncut chip thickness where first evidence of material detachment from the rake face is noted. Using this method, the precision of the estimated $h_{min}$ value is dependent on the discretization of the $h/r_n$ parameters in simulation series [25]. Furthermore Ducobu [26] noted that in the ploughing regime, while no chip is formed at first, the wedge of material in front of the tool increases in size, as the tool progresses, to eventually initiate a chip formation. Rao et al. [27] used the element deletion criterion for the Johnson-Cook fracture model in pure Lagrangian FE simulations to estimate $h_{min}$. Uncut chip thickness was modelled as a single layer of elements which were deleted once the fracture criteria were satisfied. If the selected uncut chip thickness $h$ parameter was below $h_{min}$ then elements would not be deleted but would be simply deformed and pushed down by the tool. Another well researched method for estimation of $h_{min}$ relates to the formerly mentioned principle of stagnation point on the rounded cutting edge. Other studies involving numerical simulations, either with the molecular dynamics (MD) or finite element methods (FEM) determined that stagnation of the material occurs as a wedge-shaped region, adjacent to the cutting edge, rather than as a single point. Inside the stagnation region material flow rate is close to zero and its shape and location can be found through analysis of velocity or displacement fields. Lai et al. [28] investigated nanometric cutting of copper with 3D MD simulation and through distribution of vertical displacements in atom layers a stagnation region location was found. Once the location of the stagnant area was established a stagnation angle and $h_{min}$ were geometrically determined in relation to the center of the radius of cutting edge. Hosseini and Vahdati [29] with MD simulations evaluated the effect of cutting edge radius $r_n$ on contact zone in nano-machining where they concluded that for selected cutting edge radii, the stagnant angle remained relatively constant within the investigated range of uncut chip thickness.

The presented state-of-the-art reveals that issues concerning the estimation of ploughing forces and $h_{min}$ have been undertaken in numerous studies. Nevertheless, the majority of them refer to a simple orthogonal cutting processes, as well as the turning and end milling conducted with a tools equipped with rectilinear cutting edges. On the other hand, there are still some research gaps concerning the following aspects:

- Reliable assessment of $h_{min}$ during cutting with variable uncut chip thickness and tools equipped with a circular cutting edges (e.g. ball-end mills). During ball-end milling, the variations of tool effective diameter and cutting speed occur in the contact length in the tool-workpiece interface, affecting the thermomechanical and chip decohesion phenomena.
- Modeling/measuring the ploughing forces during ball-end milling process. In this case, the evaluation of ploughing forces requires the determination of a ploughing area, based on novel analytic expressions, different than those applied in turning or end milling.
- Characterization of $h_{min}$ and ploughing forces as a function of machined surface inclinations and progressing tool wear. Machined surface inclinations and tool wear are the key factors affecting the finishing ball-end milling of a curvilinear surfaces.

Therefore – in order to fill these research gaps – this study focuses on the evaluation of ploughing phenomenon during precise ball-end milling with various machined surface inclinations. As part of the study the original model of $h_{\min}$ based on the definition of stagnation point, as well as equivalent uncut chip thickness and intended to a ball-end milling is proposed. Moreover, the novel ploughing force model based on a measured milling forces and analytical equations of geometrical elements of cut and ploughing area during ball-end milling is formulated.

Ploughing appearing during machining processes – as the contact mechanics phenomenon, contributing to the abrasion wear of workpiece and the quality of a machined surface finish – is an important problem of a tribology science. Thus, the research results of this study can lead to a better understanding of a ploughing phenomenon occurring during precise ball-end milling of a complex surfaces and consequently effective selection of process input parameters in order to obtain the machined surface finish with a high quality and operational properties.

**Nomenclature**

| | |
|---|---|
| $A\alpha$ – flank face of a tool; | $K_{tc}$, $K_{rc}$, $K_{ac}$ – tangential, radial and binormal proportionality coefficients related to shearing; |
| $A_{Dz}(h_{\min})$ – area of cut per tooth corresponding to the $h_{\min}$; | $K_{te}$, $K_{re}$, $K_{ae}$ – tangential, radial and binormal proportionality coefficients related to edge effects; |
| $A_{Dz}(\varphi)$ – cross-section of the area of cut per tooth; | $K_{tp}$, $K_{rp}$, $K_{ap}$ – tangential, radial and binormal proportionality coefficients related to ploughing; |
| $A_{Dz}(\varphi_{av})$ – area of cut corresponding to the average working angle of the tool; | $l$ – active length of the cutting edge; |
| $A_{Dze}(\varphi)$ – equivalent cross-section of the area of cut per tooth; | $l(\varphi_{av})$ – cutting edge active length corresponding to the average working angle of the tool; |
| $a_p$ – axial cutting depth; | $R$ – radius of tool; |
| $a_p(\varphi)$ – instantaneous depth of cut; | $Rz$ - maximum height of surface roughness profile |
| $a_p(\varphi_{av})$ – average depth of cut; | $r(\psi_l)$ – vector radius corresponding to the lag angle; |
| $A_{pl}$ – ploughing area; | $r_{av}$ – distance in the Z-axis of the center of gravity of the area of cut from the point P on the cutting edge; |
| $a_{tn}$, $a_{rn}$ –experimentally determined slope coefficients of regression equations of tangential and radial force | $r_{av}$ – radius vector of the center of gravity of the area of cut; |
| $A\gamma$ – rake face of a tool; | $VB$ – flank wear on the cutting edge of a ball-end mill; |
| $b$ – uncut chip width; | $VB$ – tool flank wear; |
| $b_e$ – equivalent uncut chip width; | $v_c$ – cutting speed; |
| $b_r$ – pick feed; | $v_f$ – feed speed; |
| $D$ – diameter of tool; | $V_{pl}$ – ploughing volume; |
| $e_r$ – run-out value; | $V_{pl}(h_{\min})$ – ploughing volume corresponding to the $h_{\min}$ (maximum ploughing volume); |
| $F_t$, $F_r$, $F_a$ – tangential, radial and binormal forces; | $z$ – number of teeth; |
| $F_{t\_av}$, $F_{r\_av}$, $F_{a\_av}$ – average tangential, radial and binormal forces in tool coordinate system; | $\alpha$ – surface inclination angle; |
| $F_{tc}(t)$, $F_{rc}(t)$, $F_{ac}(t)$ – instant shearing forces, tangential, radial, binormal; | $\alpha_o$ – orthogonal flank angle; |
| $F_{te}(t)$, $F_{re}(t)$, $F_{ae}(t)$ – instant edge forces, tangential, radial, binormal; | $\gamma_o$ – orthogonal rake angle; |
| $F_{te}$, $F_{re}$ – tangential and radial edge forces; | $\delta$ – run-out angle; |
| $F_{tn\_av}$, $F_{rn\_av}$ – arithmetic mean tangential force and radial force, defined in the normal plane; | $\Delta e$ – variation of area of cut induced by radial run-out; |
| $F_{tp}(t)$, $F_{rp}(t)$, $F_{ap}(t)$ – instant ploughing forces, tangential, radial, binormal; | $\Delta s$ – elastic springback of a workpiece; |
| $F_{x\_\max}$, $F_{y\_\max}$, $F_{z\_\max}$ – maximum measured values of milling forces in machine tool coordinates, namely: feed normal, feed and thrust force; | $\lambda_s$ – nominal helix angle of a ball-end mill; |
| | $\lambda_{s\_av}$ – average helix angle of a ball-end mill; |
| $f_z$ – feed per tooth; | $\varphi$ – instantaneous working angle of the tool; |
| $h$ – uncut chip thickness; | $\varphi_{av}$ – average working angle of the tool; |
| $h_{\min}$ – minimum uncut chip thickness; | $\varphi_p$ – contact angle corresponding to switching from shearing to ploughing; |
| $h_z$ –uncut chip thickness per tooth | $\varphi_r$ – cutting edge fragment position angle; |
| $h_{ze}(\varphi)$ – equivalent uncut chip thickness; | $\varphi_{r2}$ – cutting edge fragment final position angle; |
| $h_{ze\_av}$ – average equivalent uncut chip thickness per 1 tooth of the ball-end mill; | $\varphi_{re}(\varphi)$ – instantaneous equivalent position angle of the cutting edge fragment; |
| $k$ – normalized minimum uncut chip thickness; | $\psi_{l1}$ – initial lag angle; |
| | $\psi_{l2}$ – final lag angle; |
| | $\psi_n$ – working angle in a normal plane of a tool; |
| | $\Omega$ – tool rotation angle; |

## 2. ANALYTIC-EXPERIMENTAL MODEL OF PRECISE BALL-END MILLING

### 2.1. General formulation of model

In this chapter, an attempt is made to formulate a generalized model of the cutting force components, and $h_{min}$ concerning precise cutting with ball-end mills. For this purpose, the approaches proposed by the authors of works [30, 31] were adapted. Original models for the geometric parameters of the area of cut and proportionality coefficients were formulated as part of own research. These models take into account the ploughing effect of the material, and the immersion boundaries of tool in the workpiece material, which are characteristic for micromilling and ball-end milling of curvilinear surfaces.

The intensity of the phenomena characteristic for a precision machining, affecting the generated forces, depends on the relationship between the uncut chip thickness and its minimum value $h_{min}$. In general, during machining in the range of $f_z \gg h_{min}$, the forces acting on the tool ($F_t$, $F_r$, $F_a$) depend on the shear phenomenon appearing in the tool rake face-chip interface, as well as on a friction between the tool flank face and the machined material. When $f_z \approx h_{min}$, the forces generated during cutting also depend on the phenomenon of ploughing, characteristic for a machining in the range of $h < h_{min}$ (Fig. 1). Another important factor determining the forces during precision machining is a tool wear. Occurrence of abrasion on the tool flank face $VB$ increases the contact length between this surface and the machined workpiece and thus affects the values of forces.

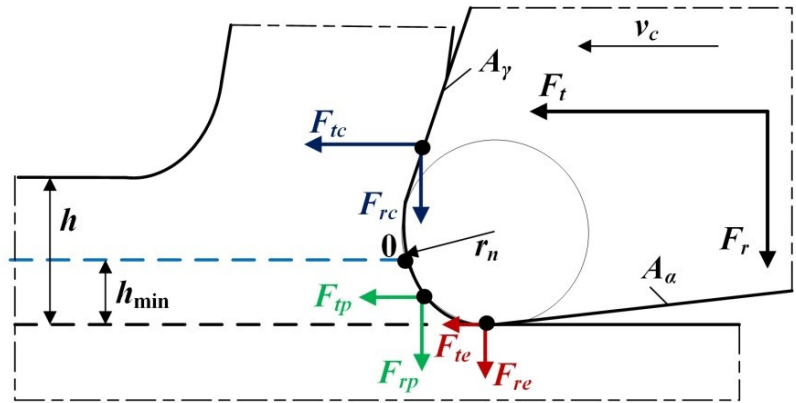

**Fig. 1.** Forces on the rounded cutting edge of the tool for $f_z \approx h_{min}$ (using orthogonal cutting as an example)

It should be noted that the separation boundary between cutting and intensive ploughing is difficult to define unambiguously. It has been conventionally proposed to identify the occurrence of a cutting area with intensive ploughing for $h_{min}/f_z > 0.1$. This is motivated by the relationship between the area of decohesion of the material and the ploughing area, per one cycle of tool penetration into the wokrpiece. During slot milling, when $h_{min}/f_z < 0.1$, the contact angle $\varphi_p$, corresponding to the switching from ploughing to shearing, is less than 10° (Figure 2).

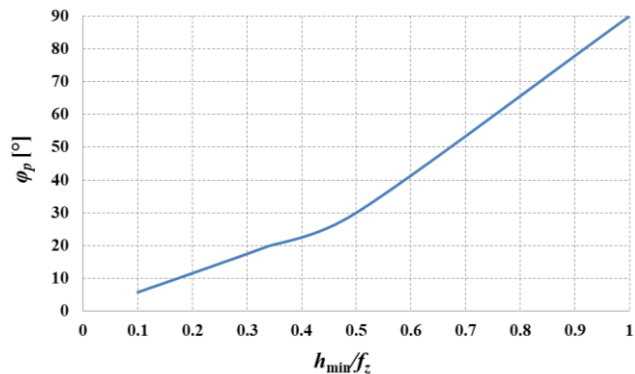

**Fig. 2.** Influence of the ratio of $f_z$ relative to the $h_{min}$ on the contact angle corresponding to switching from ploughing to shearing during slot milling

## 2.2. Model of minimum uncut chip thickness for a ball-end milling process

As part of this section, the $h_{min}$ was assessed on the basis of the stagnant point 0 identification method for the end milling process [5] and extended to the ball-end mills. The stagnation point defined on the workpiece-rounded cutting edge interface (see – Fig. 1) determines the changes in workpiece material flow direction. Therefore, the material of workpiece, which moves above the stagnant point transforms to the chip. On the other side, the workpiece material which flows below the stagnant point is being pushed under the flank face of the tool (no chip formation occurs). In this method, the location of a stagnant point determines the direction change of a force tangent to the curved cutting edge.

The general form of equation for the determination of $h_{min}$ is formulated in the following form:

$$h_{min} = r_n \left[ 1 - \sin\left( \arctan\left( \frac{a_{rn}}{a_{tn}} \right) \right) \right] \quad (1)$$

Equation (1) reveals that estimating the $h_{min}$ requires knowing the values of the slope coefficients $a_{tn}$, $a_{rn}$ of the regression equations of tangential and normal force, and the value of the cutting edge radius $r_n$. Therefore, in order to apply this method, experimental or numerical studies are required, including measurements of a cutting force components in function of the feed per tooth and the microgeometry of the cutting edge.

The regression models of the arithmetic mean tangential force and radial force, defined in the normal plane ($F_{tn\_av}$, $F_{rn\_av}$), are described by the equations:

$$\begin{aligned} F_{tn\_av} &= a_{tn} \cdot h_{ze\_av} + F_{te} \\ F_{rn\_av} &= a_{rn} \cdot h_{ze\_av} + F_{re} \end{aligned} \quad (2)$$

where:
$a_{tn}$, $a_{rn}$ – experimentally determined slope coefficients of regression equations of tangential and radial force
$h_{ze\_av}$ – average equivalent uncut chip thickness per 1 tooth of the ball-end mill
$F_{te}$, $F_{re}$ – tangential and radial edge forces.

Determining the average tangential force in the normal plane and radial force in the normal plane of a ball-end mill, necessary for estimating the $h_{min}$, requires taking into consideration more factors than when determining the values of these components in turning, or cylindrical-face milling. The reason for this is the influence of the instantaneous working angle $\varphi$, diameter of tool $D$, axial cutting depth $a_p$, and the machined surface inclination angle $\alpha$ on the relationship between the forces in the tool coordinate system and force components in the machine tool system.

The first step is to determine the average equivalent uncut chip thickness per 1 tooth of the ball-end mill $h_{ze\_av}$, taking into account the curved geometry of the instantaneous area of the cut (Fig. 3). The curvilinear profile of the area of the cut implies a variation in the value of the uncut chip thickness as a function of the position angle of the cutting edge fragment $\varphi_r$ regardless of the considered working angle $\varphi$. It substantially hinders the evaluation of forces distributed onto the area of cut, and thus prediction of the $h_{min}$ value. In order to simplify the issue of the uncut chip thickness during machining with a spherical cutter, it was assumed that the value of the force acting on any cross-section of the area of cut $A_{Dz}(\varphi)$ does not depend on its shape, but only on its value, and that the forces are applied at the center of gravity of this field. This made it possible to introduce the so-called equivalent uncut chip thickness $h_{ze}(\varphi)$ acting on a rectangular equivalent cross-section of the area of cut $A_{Dze}(\varphi)$, satisfying the condition: $A_{Dze}(\varphi) = A_{Dz}(\varphi)$ (Fig. 3). This condition assumes that rectangular equivalent cross-section of the area of cut $A_{Dze}(\varphi)$ and cross-section of the area of cut $A_{Dz}(\varphi)$ differ only in terms of shape, but have the same values. This approach is consistent with equations describing milling force models applied in machining science (e.g. models developed by Park and Malekian [30], and Zhang et al. [31]). These models assume that shearing forces are expressed by the product of specific force coefficients and area of cut (which in turn is the product of uncut chip thickness and uncut chip width).

Therefore, values of forces are directly related to the values of area of cut, instead of its shape. Moreover in these models, the elemental force vectors are located in the center of gravity of the elemental areas of cut.

On the basis of above assumptions, the value of the equivalent uncut chip thickness per 1 tooth $h_{ze}(\varphi)$ can be determined for any instantaneous tool operating angle, using the equation:

$$h_{ze}(\varphi) = R + f_z \cdot \sin \varphi_{re}(\varphi) - \sqrt{R^2 - f_z^2 \cdot \cos \varphi_{re}(\varphi)} \tag{3}$$

The instantaneous equivalent position angle of the cutting edge fragment, which can be found in equation (3), is determined based on expression (4):

$$\varphi_{re}(\varphi) = \frac{1}{3}\left[\arctan\left(\frac{\sqrt{a_p(\varphi) \cdot (D - a_p(\varphi))}}{R - a_p(\varphi)}\right) + \arctan\left(\frac{\sqrt{a_p(\varphi) \cdot (D - a_p(\varphi))} - f_z \cdot \sin \varphi}{R - a_p(\varphi)}\right)\right] \tag{4}$$

In order to calculate a value of the average (equivalent) uncut chip thickness per 1 tooth $h_{ze\_av}$, it is necessary to take into the consideration the value of the average working angle of the tool $\varphi_{av}$ in equations (3, 4). A value of $\varphi_{av}$ angle is dependent on the conditions of mapping the tool in the workpiece material, related to the surface inclination angle (Fig. 4). In principle, two basic kinematic variations of machining with a spherical cutter can be distinguished: without inclination of the tool axis/machined surface (slot milling) – $\alpha = 0$ (Fig. 4a) and with inclination (ramping) – $\alpha > 0$ (Fig. 4b).

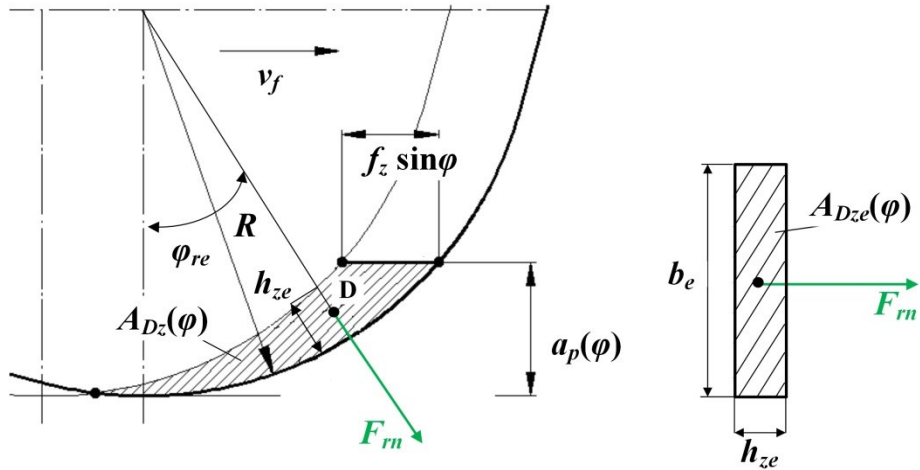

**Fig. 3.** Area of cut and equivalent area of cut in ball-end milling

For symmetric milling with $\alpha = 0$ it is assumed that $\varphi_{av} = \pi/4$, and $a_p(\varphi_{av}) = a_p$. In symmetrical milling with $\alpha > 0$ the value of the average tool working angle $\varphi_{av}$ is determined by the equation:

$$\varphi_{av} = \frac{\pi}{2} - \frac{1}{2}\left[\arccos\left(1 - \frac{a_p}{R \cdot \sin^2 \alpha}\right)\right] \tag{5}$$

In order to calculate the average depth of cut for the case when $\alpha > 0$, the following equation is formulated:

$$a_p(\varphi_{av}) = R - \left((R - a_p) \cdot \cos\left[\arctan\left(\operatorname{ctg} \varphi_{av}\right) \cdot \sin \alpha\right]^{-1}\right) \tag{6}$$

Determining the values of the $\varphi_{av}$ and $\varphi_{re}(\varphi_{av})$ angles enables calculation of forces $F_{tn\_av}$ and $F_{rn\_av}$. For this purpose, for the angle $\alpha = 0$, the equations were adapted:

$$F_{tn\_av} = \frac{\sqrt{2}}{2} \begin{bmatrix} (F_{x\max} \cdot \sin\varphi_{av} - F_{y\max} \cdot \cos\varphi_{av})\cos\lambda_{s\_av} + \\ + \begin{pmatrix} -F_{x\max} \cdot \cos\varphi_{av} \cdot \cos\varphi_{re}(\varphi_{av}) - \\ -F_{y\max} \cdot \sin\varphi_{av} \cdot \cos\varphi_{re}(\varphi_{av}) - \\ -F_{z\max} \cdot \sin\varphi_{re}(\varphi_{av}) \end{pmatrix} \sin\lambda_{s\_av} \end{bmatrix} \quad (7)$$

$$F_{rn\_av} = \frac{\sqrt{2}}{2} \begin{bmatrix} -F_{x\max} \cdot \cos\varphi_{av} \cdot \sin\varphi_{re}(\varphi_{av}) + F_{z\max} \cdot \cos\varphi_{re}(\varphi_{av}) - \\ -F_{y\max} \cdot \sin\varphi_{av} \cdot \sin\varphi_{re}(\varphi_{av}) \end{bmatrix} \quad (8)$$

where:

$F_{x\_\max}$, $F_{y\_\max}$, $F_{z\_\max}$ – maximum measured values of milling forces in machine tool coordinates, namely: feed normal, feed and thrust force.

a)

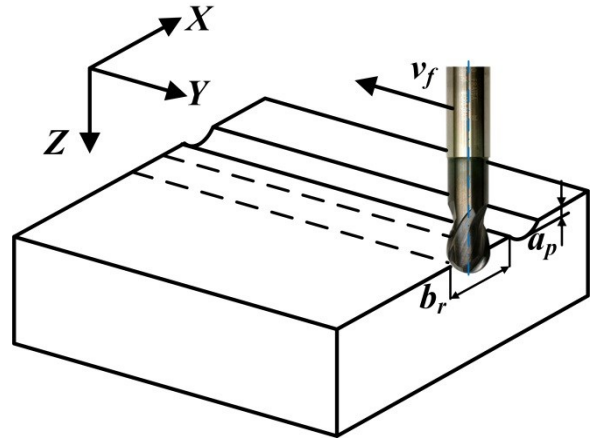

b)

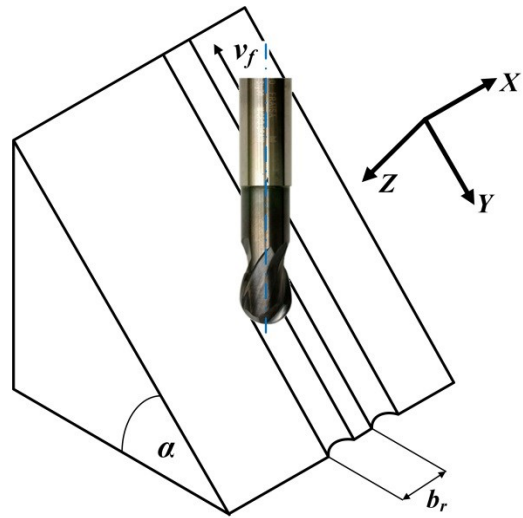

**Fig. 4.** The scheme of symmetrical ball-end milling with: a) $\alpha = 0$, b) $\alpha > 0$

For a ball-end milling conducted with an inclined machined surface ($\alpha > 0$) the forces $F_{tn\_av}$ and $F_{rn\_av}$ are expressed by the equations:

$$F_{tn\_av} = \frac{1}{2} \begin{bmatrix} (F_{x\max} \cdot \sin\varphi_{av} - F_{y\max} \cdot \cos\varphi_{av})\cos\lambda_{s\_av} + \\ + \begin{pmatrix} -F_{x\max} \cdot \cos\varphi_{av} \cdot \cos\varphi_{re}(\varphi_{av}) - \\ -F_{y\max} \cdot \sin\varphi_{av} \cdot \cos\varphi_{re}(\varphi_{av}) - \\ -F_{z\max} \cdot \sin\varphi_{re}(\varphi_{av}) \end{pmatrix} \sin\lambda_{s\_av} \end{bmatrix} \quad (9)$$

$$F_{rn\_av} = \frac{1}{2}\left[\begin{array}{l}-F_{x\max}\cdot\cos\varphi_{av}\cdot\sin\varphi_{re}(\varphi_{av})+F_{z\max}\cdot\cos\varphi_{re}(\varphi_{av})-\\-F_{y\max}\cdot\sin\varphi_{av}\cdot\sin\varphi_{re}(\varphi_{av})\end{array}\right] \quad (10)$$

During machining with a ball-end milling tool equipped with a helical profile of the cutting edge ($\lambda_s \neq 0$), the value of the angle $\lambda_s$ is variable on a section of the cutting edge, located in the spherical region of the tool working part. Therefore, in the proposed force models $F_{tn\_av}$, $F_{rn\_av}$, the average value of the $\lambda_{s\_av}$ angle, associated with the point on the cutting edge that defines the center of gravity of the area of the cut for the angle $\varphi_{av}$ is considered. The value of the $\lambda_{s\_av}$ angle is described by the expression:

$$\lambda_{s\_av} = \arctan\left(\frac{r_{av}\cdot\mathrm{tg}\,\lambda_s}{R}\right) \quad (11)$$

The parameter $r_{av}$ in equation (11) defines the radius vector of the center of gravity of the area of cut and is expressed by the equation:

$$r_{av} = \sqrt{R^2 - (R - z_{av})^2} \quad (12)$$

Determining the value of the $r_{av}$ parameter from equation (12) requires calculating the distance in the Z-axis of the center of gravity of the area of cut from the point P on the cutting edge, according to the relations:

$$z_{av} = R\left(1 - \cos\varphi_{re}(\varphi_{av})\right) \text{ for } \alpha = 0 \quad (13)$$

$$z_{av} = R\left(1 - \cos\left\{\frac{1}{2}\left[2\alpha + \arccos\left(\frac{R - a_p}{R}\right)\right]\right\}\right) \text{ for } \alpha > 0 \quad (14)$$

Determined values of the $F_{tn\_av}$, $F_{rn\_av}$ components and equivalent uncut chip thickness $h_{ze\_av}$ allow the formulation of the force model described by equation (2), and subsequently the prediction of the $h_{\min}$ according to equation (1). However, in order to characterize the effect of surface inclination angle $\alpha$ on the minimum uncut chip thickness $h_{\min}$, the $F_{tn\_av}$, $F_{rn\_av} = f(h_{ze\_av})$ regression models have to be determined for various tested $\alpha$ angles, and then the $h_{\min}$ values for corresponding surface inclination angles are being calculated.

### 2.3. Shearing, edge and ploughing force model

Based on the studies [30, 31] and the relationships shown in figure 1, the general form of the components of a total cutting force acting on a cutting edge was formulated:

$$\begin{cases} F_t(t) = F_{tc}(t) + F_{te}(t) \\ F_r(t) = F_{rc}(t) + F_{re}(t), \quad h_{\min}/f_z \leq 0.1 \\ F_a(t) = F_{ac}(t) + F_{ae}(t) \end{cases} \quad (15)$$

$$\begin{cases} F_t(t) = F_{tc}(t) + F_{te}(t) + F_{tp}(t) \\ F_r(t) = F_{rc}(t) + F_{re}(t) + F_{rp}(t), \quad h_{\min}/f_z > 0.1 \\ F_a(t) = F_{ac}(t) + F_{ae}(t) + F_{ap}(t) \end{cases} \quad (16)$$

where:
$F_{tc}(t)$, $F_{rc}(t)$, $F_{ac}(t)$ – instant shearing forces, tangential, radial, binormal
$F_{te}(t)$, $F_{re}(t)$, $F_{ae}(t)$ – instant edge forces, tangential, radial, binormal
$F_{tp}(t)$, $F_{rp}(t)$, $F_{ap}(t)$ – instant ploughing forces, tangential, radial, binormal

The equations can be used to determine the instantaneous shearing forces:

$$\begin{cases} F_{tc}(t) = K_{tc} \cdot A_{Dz}(t) \\ F_{rc}(t) = K_{rc} \cdot A_{Dz}(t) \\ F_{ac}(t) = K_{ac} \cdot A_{Dz}(t) \end{cases} \quad (17)$$

Instantaneous edge forces are expressed by the formulas:

$$\begin{cases} F_{te}(t) = K_{te} \cdot l(t) \\ F_{re}(t) = K_{re} \cdot l(t) \\ F_{ae}(t) = K_{ae} \cdot l(t) \end{cases} \quad (18)$$

Instantaneous values of ploughing forces can be determined using the following equations:

$$\begin{cases} F_{tp}(t) = K_{tp} \cdot V_{pl}(t) \\ F_{rp}(t) = K_{rp} \cdot V_{pl}(t) \\ F_{ap}(t) = K_{ap} \cdot V_{pl}(t) \end{cases} \quad (19)$$

Equations (17-19) reveal that the estimation of a total cutting force components values during precision cutting requires knowledge of the proportionality coefficients related to shearing ($K_{tc}$, $K_{rc}$, $K_{ac}$), edge ($K_{te}$, $K_{re}$, $K_{ae}$) and ploughing ($K_{tp}$, $K_{rp}$, $K_{ap}$) effects, as well as the geometric parameters of the cut (area of cut – $A_{Dz}$, active length of the cutting edge $l$ and ploughing volume $V_{pl}$).

In the case of a spherical cutter, the area of the cut taking into account radial run-out is expressed by the relation:

$$A_{Dz}(\varphi) = b \cdot h_z(\varphi, \varphi_r) = R \cdot (f_z + \Delta e) \cdot (1 - \cos(\varphi_{r2} - \alpha)) \cdot \sin \varphi \quad (20)$$

where:
$\varphi_r$ – cutting edge fragment position angle
$\varphi_{r2}$ – cutting edge fragment final position angle.

The value of the parameter $\Delta e$ for a spherical cutter can be determined from the equation:

$$\Delta e = e_r \cdot \sin\left(\Omega - \frac{\psi_{l1} + \psi_{l2}}{2} + \delta\right) \quad (21)$$

where:
$\psi_{l1}$ – initial lag angle
$\psi_{l2}$ – final lag angle.

To determine the cutting edge active length for a spherical milling tool, the following equation can be applied:

$$l = \int_{\psi_{l1}}^{\psi_{l2}} \sqrt{\left(\frac{dr(\psi_l)}{d\psi_l}\right)^2 + r(\psi_l)^2 + \frac{R^2}{\mathrm{tg}^2 \lambda_s}} \, d\psi_l \quad (22)$$

The vector radius corresponding to the lag angle $r(\psi_l)$, contained in equation (22), is expressed by:

$$r(\psi_l) = R\sqrt{1 - \left(\frac{\psi_l}{\mathrm{tg}\,\lambda_s} - 1\right)^2} \quad (23)$$

Determining the values of $A_{Dz}$ and $l$ from equations (20-23) requires defining the boundary conditions for tool immersion in the workpiece material, described (in the case of ball-end milling cutters) by angles $\psi_{l1}$, $\psi_{l2}$, $\varphi_{r1}$, $\varphi_{r2}$, $\psi$. The values of these angles are strictly dependent on the kinematics and strategy of ball-end milling. The details concerning the calculation of a boundary conditions ($\psi_{l1}$, $\psi_{l2}$, $\varphi_{r1}$, $\varphi_{r2}$, $\psi$) during ball-end milling with various kinematics are presented in a previous works [32, 33].

The next essential factor which should be considered in a proposed model is a ploughing volume. The ploughing volume mainly depends on a degree of elastic and plastic deformation of the material flowing onto the cutting edge (expressed by the values of $h_{min}$ and elastic springback $\Delta s$), as well as the cutting edge microgeometry and instantaneous value of the uncut chip thickness ($h < h_{min}$ – Fig. 5).

Based on the trigonometric relationships shown in figure 5, an expression for the ploughing area $A_{pl}$ can be formulated:

$$A_{pl} = \frac{1}{2}\left(\psi_n \cdot r_n^2 - r_n \sin\psi_n \cdot (r_n - h) + \frac{\Delta_s^2}{\text{tg}^2\alpha_n}\right) \tag{24}$$

The angle $\psi_n$ in equation (24) is determined from the following equation:

$$\psi_n = \arccos\left(1 - \frac{h}{r_n}\right) \tag{25}$$

To determine the ploughing volume, the following equation should be applied:

$$V_{pl} = A_{pl} \cdot b \tag{26}$$

For a spherical cutter, it is assumed that uncut chip width corresponds to the value of the equivalent uncut chip width, i.e. $b = b_e$. The $b_e$ parameter value is calculated from the equation:

$$b_e = \frac{A_{Dz}(\varphi)}{h_{ze}(\varphi)} \tag{27}$$

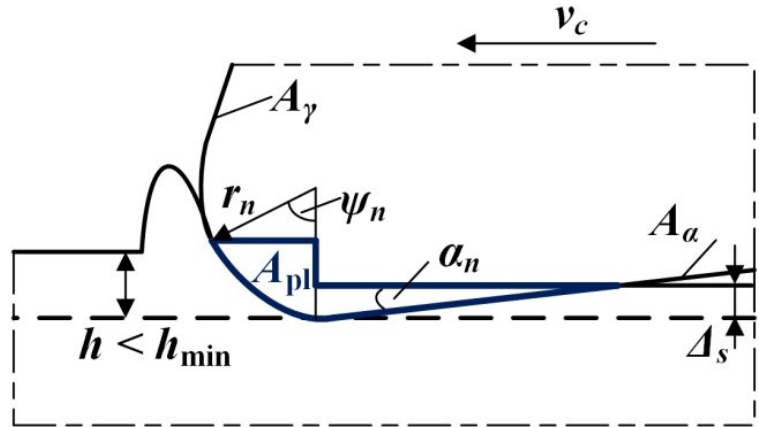

**Fig. 5.** Model of the ploughing area

The next step of force prediction based on the proposed approach is to determine the values of proportionality coefficients related to shearing ($K_{tc}$, $K_{rc}$, $K_{ac}$), edge ($K_{te}$, $K_{re}$, $K_{ae}$) and ploughing ($K_{tp}$, $K_{rp}$, $K_{ap}$) effects. The method presented here is based on determining the average values of forces in the tool coordinate system ($F_{t\_av}$, $F_{r\_av}$, $F_{a\_av}$) during milling tests carried out for different values of surface inclination angles.

To determine the average values of forces in tool coordinate system, a measured maximum values of the forces in the machine tool coordinate system ($F_{xmax}$, $F_{ymax}$, $F_{zmax}$) must be transformed. In the case of full immersion milling with a ball-end mill and angle $\alpha = 0$, the values of the forces $F_{t\_av}$, $F_{r\_av}$, $F_{a\_av}$ can be expressed by the equations:

$$F_{t\_av} = \frac{\sqrt{2}}{2}\left(F_{xmax} \cdot \sin\varphi_{av} - F_{ymax} \cdot \cos\varphi_{av}\right) \tag{28}$$

$$F_{r\_av} = \frac{\sqrt{2}}{2}\left[\begin{pmatrix}-F_{x\max}\cdot\cos\varphi_{av}\cdot\sin\varphi_{re}(\varphi_{av})+F_{z\max}\cdot\cos\varphi_{re}(\varphi_{av})-\\-F_{y\max}\cdot\sin\varphi_{av}\cdot\sin\varphi_{re}(\varphi_{av})\end{pmatrix}\right] \quad (29)$$

$$F_{a\_av} = \frac{\sqrt{2}}{2}\left[\begin{pmatrix}-F_{x\max}\cdot\cos\varphi_{av}\cdot\cos\varphi_{re}(\varphi_{av})-F_{z\max}\cdot\sin\varphi_{re}(\varphi_{av})-\\-F_{y\max}\cdot\sin\varphi_{av}\cdot\cos\varphi_{re}(\varphi_{av})\end{pmatrix}\right] \quad (30)$$

In case of a ball-end milling of an inclined machined surface ($\alpha \neq 0$), the average forces in tool coordinate system are expressed by the following relationships:

$$F_{t\_av} = \frac{1}{2}\left(F_{x\max}\cdot\sin\varphi_{av} - F_{y\max}\cdot\cos\varphi_{av}\right) \quad (31)$$

$$F_{r\_av} = \frac{1}{2}\left[\begin{pmatrix}-F_{x\max}\cdot\cos\varphi_{av}\cdot\sin\varphi_{re}(\varphi_{av})+F_{z\max}\cdot\cos\varphi_{re}(\varphi_{av})-\\-F_{y\max}\cdot\sin\varphi_{av}\cdot\sin\varphi_{re}(\varphi_{av})\end{pmatrix}\right] \quad (32)$$

$$F_{a\_av} = \frac{1}{2}\left[\begin{pmatrix}-F_{x\max}\cdot\cos\varphi_{av}\cdot\cos\varphi_{re}(\varphi_{av})-F_{z\max}\cdot\sin\varphi_{re}(\varphi_{av})-\\-F_{y\max}\cdot\sin\varphi_{av}\cdot\cos\varphi_{re}(\varphi_{av})\end{pmatrix}\right] \quad (33)$$

To determine the edge forces ($F_{te}$, $F_{re}$, $F_{ae}$) it is necessary to formulate linear regression equations of the components $F_{t\_av}$, $F_{r\_av}$, $F_{a\_av} = f(h_{ze\_av})$. The constant terms in the resulting equations are the edge forces (see equation (2)).

The resulting force values ($F_{te}$, $F_{re}$, $F_{ae}$, $F_{t\_av}$, $F_{r\_av}$, $F_{a\_av}$) allow the calculation of edge and shear-related coefficients according to the equations:

$$K_{te} = \frac{F_{te}}{l(\varphi_{av})}, \quad K_{re} = \frac{F_{re}}{l(\varphi_{av})}, \quad K_{ae} = \frac{F_{ae}}{l(\varphi_{av})} \quad (34)$$

$$K_{tc} = \frac{F_{t\_av} - F_{te}}{A_{Dz}(\varphi_{av})}, \quad K_{rc} = \frac{F_{r\_av} - F_{re}}{A_{Dz}(\varphi_{av})}, \quad K_{ac} = \frac{F_{a\_av} - F_{ae}}{A_{Dz}(\varphi_{av})} \quad (35)$$

where:

$l(\varphi_{av})$ – cutting edge active length corresponding to the average working angle of the tool

$A_{Dz}(\varphi_{av})$ – area of cut corresponding to the average working angle of the tool.

For the purpose of a ploughing coefficients determination, it was assumed that the maximum value of the ploughing force per tooth corresponds to the value of the minimum uncut chip thickness. Subsequently, the parameters $K_{tp}$, $K_{rp}$, $K_{ap}$ can be expressed by equations:

$$K_{tp} = \frac{K_{tc}\cdot A_{Dz}(h_{\min})}{V_{pl}(h_{\min})}, \quad K_{rp} = \frac{K_{rc}\cdot A_{Dz}(h_{\min})}{V_{pl}(h_{\min})}, \quad K_{ap} = \frac{K_{ac}\cdot A_{Dz}(h_{\min})}{V_{pl}(h_{\min})} \quad (36)$$

where:

$A_{Dz}(h_{\min})$ – area of cut per tooth corresponding to the $h_{\min}$

$V_{pl}(h_{\min})$ – ploughing volume corresponding to the $h_{\min}$ (maximum ploughing volume).

In order to consider the influence of a surface inclination angle $\alpha$ on the edge and ploughing forces, the proportionality coefficients were calculated from equations (34 – 36) for a measured milling force data recorded during machining experiments with various $\alpha$ angles. In the next step, the proportionality coefficients regression models were expressed as a function of surface inclination angle $\alpha$ in a form of a quadratic equations. Ultimately, the proportionality coefficients regression models $K_{ic}$, $K_{ie}$, $K_{ip} = f(\alpha)$ allowed the formulation of the force model described by equations (15 and 16) for a various surface inclination angles.

## 3. EXPERIMENTAL DETAILS

The conducted research included milling tests on hardened AISI L6 (55NiCrMoV6) tool alloy steel (average hardness of 58 HRC). As part of milling experiment, values of the total cutting force components were measured to determine the forces $F_{tn\_av}$, $F_{rn\_av}$, and then to predict the $h_{min}$.

Ball-end mills applied in the experiments were TiAlN-coated solid carbide ball-end mills ($D$ = 8 mm, $z$ = 2, $\gamma_o = -10°$, $\alpha_o = 6°$, $\lambda_s = 30°$ – measured on cylindrical part of cutter). A five-axis milling center (DMU 60monoBLOCK) has been applied for a ball-end milling tests. Milling trials were conducted without the use of cutting fluids (dry cutting). Milling forces were acquired with a Kistler 9256C1 piezoelectric dynamometer in the three directions defined in the machine tool coordinates (Figure 6). The variable factors in the conducted experiments were the feed per tooth $f_z$, and the machined surface inclination angle $\alpha$ (Tab. 1). In addition, in order to obtain the constant maximum cutting speed $v_{cmax}$ during ball-end milling, with distinct surface inclinations, the variable rotational speeds were selected based on equation presented in [32]. The choice of feed rate as a variable parameter was dictated by the need to formulate the force models $F_{tn\_av}$, $F_{rn\_av}$ = f($h$) necessary to determine the value of the minimum uncut chip thickness and to estimate specific cutting force coefficients. The surface inclination angle of the machined surface is an important variable when machining curved surfaces, determining both the state of the geometrical surface specification, as well as the values of the milling forces.

As part of a further study, the microgeometries of the tested tool working parts were measured using an Alicona IF – Edgemaster optical device to obtain the cutting edge radius $r_n$. Subsequently, the tool wear of the ball-end mills was evaluated with a use of scanning electron microscope (SEM, type TESCAN Vega 3), containing a secondary electron detector. As part of measurements, the tool wear on a flank face, as well as onto the rounded cutting edge were evaluated.

In the last stage, the surface topographies of a machined surfaces were examined using an optical Veeco NT 1100 profiler, with the 5.1-fold magnification of the measured area. Moreover, in order to validate a developed minimum uncut chip thickness model, the maximum height of surface roughness profile $Rz$ (defined as the absolute vertical distance between the maximum profile peak height and the maximum profile valley depth along the sampling length) was measured in the bottom of the milled groove, in the direction parallel to the feed motion. The $Rz$ parameter was obtained applying the sampling length $lr$ = 0.8 mm, the evaluation length $ln$ = 4.0 mm, and the length of cutoff wave $\lambda c$ = 0.8 mm. The experimental set-up together with a research stages was presented in figure 6.

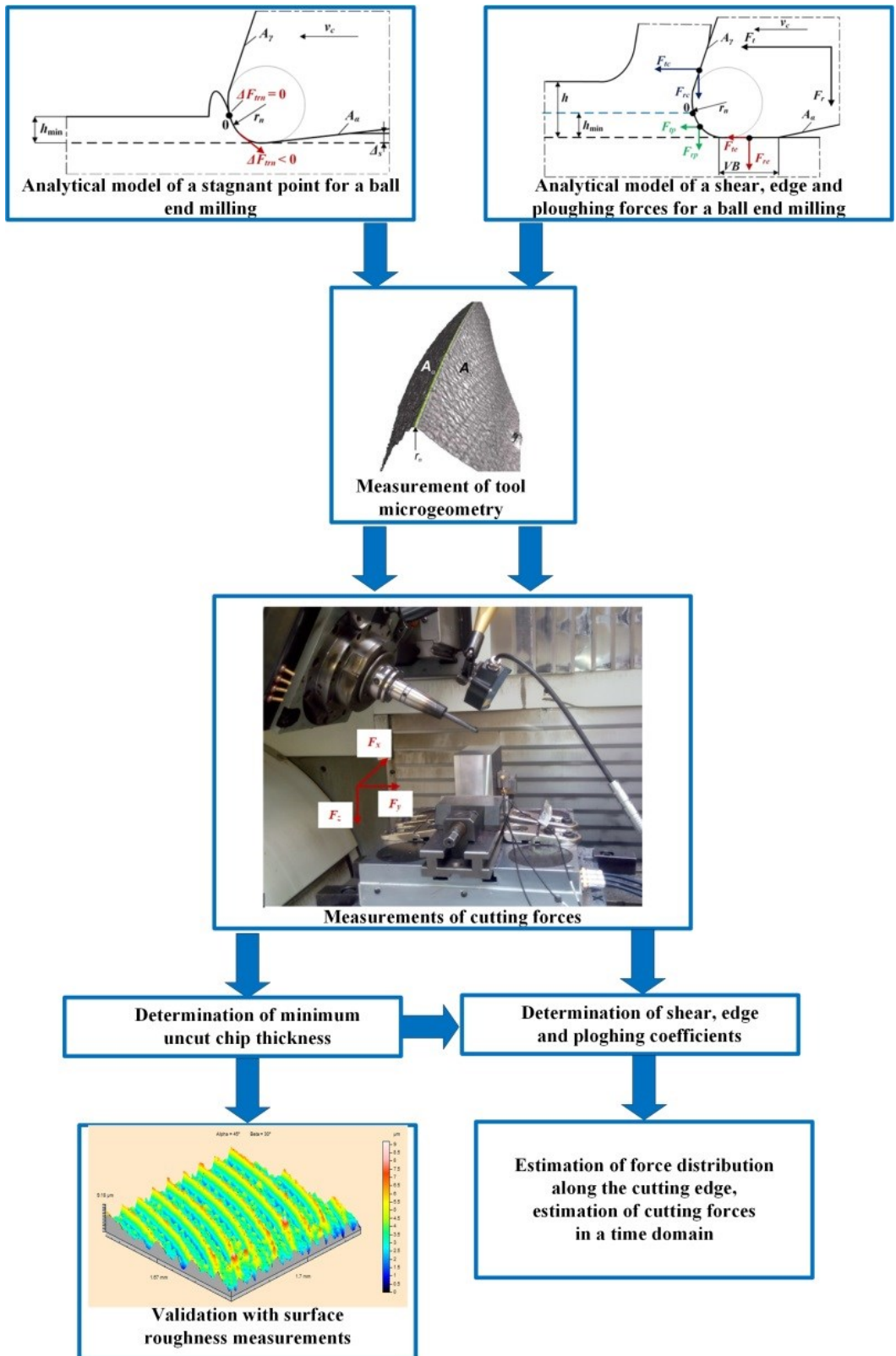

**Fig. 6.** The experimental set-up and research stages

**Tab. 1.** The applied ball-end milling conditions

| $v_{c\max}$ [m/min] | $f_z$ [mm/tooth] | $a_p$ [mm] | $\alpha$ [°] |
|---|---|---|---|
| 100 | 0.02; 0.04; 0.06; 0.08; 0.1 | 0.2 | 0; 15; 30; 45 |

## 4. RESULTS AND DISCUSSION
### 4.1. Evaluation of minimum uncut chip thickness

Figures 7 and 8 show the tangential and radial force courses in function of the $h_{ze\_av}$ parameter, and table 2 shows the regression equations of the forces $F_{tn\_av}$, $F_{rn\_av} = f(h_{ze\_av})$. The error bars depicted in figures 7 and 8 refer to the range of force values. It is observed that growth of uncut chip thickness implies an increase in the values of $F_{tn\_av}$, $F_{rn\_av}$ forces. During ball-end milling, a dependencies between the determined forces and the uncut chip thickness are described by a linear relationship. This is consistent with the trends observed in conventional milling. An indicator of the high degree of compliance between an experimental signals and the linear force model is the coefficient of determination $R^2 > 0.95$ (Tab. 2).

a)
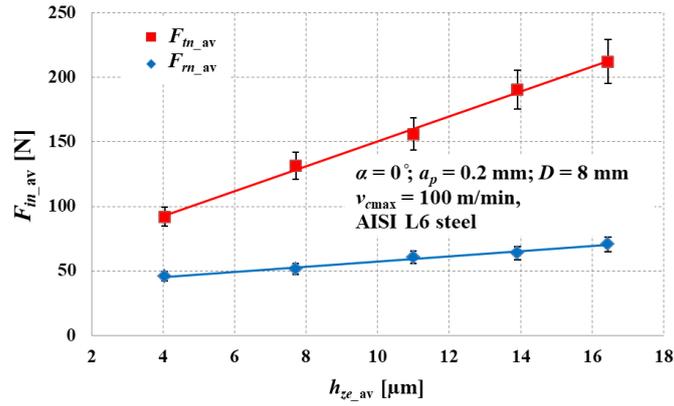

b)
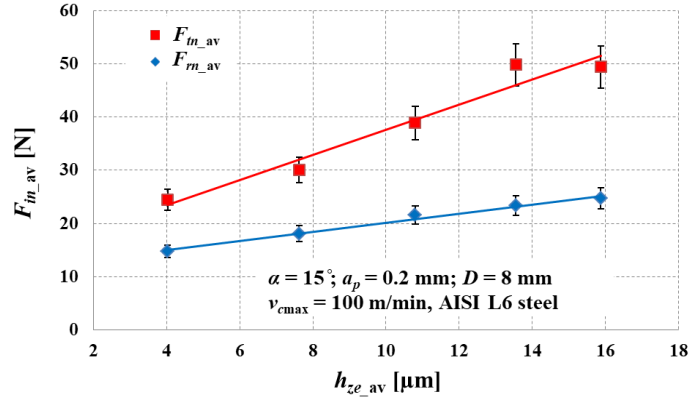

c)
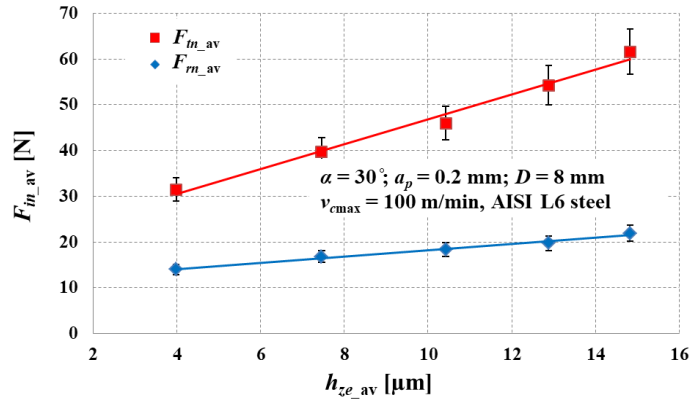

**Fig. 7.** Effect of uncut chip thickness on $F_{tn\_av}$, $F_{rn\_av}$ forces in ball-end milling with: a) $\alpha = 0$, $VB_B \approx 0$, b) $\alpha = 15°$, $VB_B \approx 0$, c) $\alpha = 30°$, $VB_B \approx 0$

a)

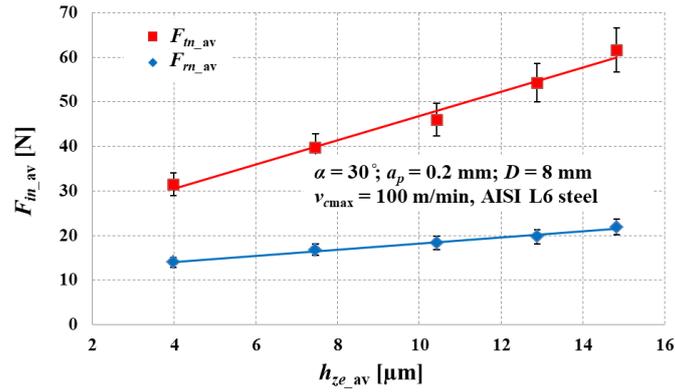

b)

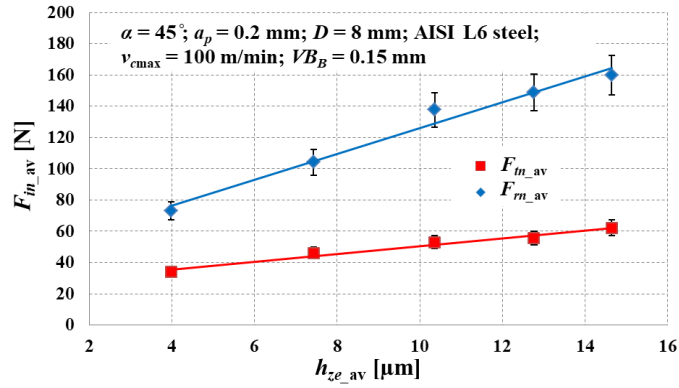

**Fig. 8.** Effect of uncut chip thickness on $F_{tn\_av}$, $F_{rn\_av}$ forces in ball-end milling with: a) $\alpha = 45°$, $VB_B \approx 0$, b) $\alpha = 45°$, $VB_B \approx 0.15$ mm

It can be observed that in case of a ball-end milling at the initial stage of tool flank wear ($VB_B \approx 0$), a values of the tangential force are greater than the radial force regardless of the $\alpha$ angle and the $h_{ze\_av}$ parameter. The vector of the tangential component $F_{tn\_av}$ is related to the vector of the cutting speed. Thus, a direction of $F_{tn\_av}$ force is directly related to the phenomenon of material decohesion, which implies relatively high values of this force compared to the values of the $F_{rn\_av}$ component. On the other hand, in the case of machining with a worn tool ($VB_B = 0.15$ mm), higher values of the $F_{rn\_av}$ component were recorded in relation to the values of the $F_{tn\_av}$ force. The radial component is directed normally to the machined surface and therefore it is sensitive to phenomena occurring in the tool flank face-workpiece interface. Progressive wear of the cutting edge (localized on a flank face) and intensive ploughing phenomena (characteristic of low-feed machining) can lead to an increase in the share of the radial force in the total cutting force. An increase in tool wear is also associated with an increase in values of $F_{tn\_av}$, $F_{rn\_av}$ forces. This is correlated directly with a change in the microgeometry of the tool (as the tool wear progresses) and an increase in the contact length between the flank face and workpiece material, contributing to an increase in the value of the frictional force on the flank face. The analysis of the tangential and radial forces also shows that the highest force values are observed in the machining process with surface inclination angle of $\alpha = 0$.

As part of further research, the microgeometry of the tested tools was measured. The measurements show that for the new tool an average value of the cutting edge radius $r_n$ is 3.5 μm (Fig. 9). The relatively low values of the $r_n$ parameter obtained from the measurements confirm the validity of using these tools for a precision machining. In the case of a worn tool ($VB_B = 0.15$ mm), the average cutting edge radius $r_n = 28$ μm. This shows that even relatively moderate wear of the cutting tool can lead to an intense increase in the value of the $r_n$ parameter. A similar findings were reached by Afazov et al. [34] in a study related to micromilling of the AISI 4340 steel with a solid carbide end mill with a diameter equal to $D = 0.5$ mm. According to the results of this study, a cutting edge radius grown from $r_n = 1.5$ μm for a new tool to $r_n = 25$ μm after the tool travelled a cutting distance of 600 mm.

Determination of the slope coefficients $a_{tn}$, $a_{rn}$ in the tangential and normal force equations (Tab. 2) and measurements of the $r_n$ value made it possible to estimate the value of $h_{min}$ and $k$ (defined as $h_{min}/r_n$ ratio) parameters in the scope of the tests carried out (Fig. 10). Considering a ball-end milling of hardened steel, one can observe a non-monotonic effect of the $\alpha$ angle on the values of $h_{min}$ and $k$ parameters. A highest values of $h_{min}$ and $k$ occur in milling with $\alpha = 0$. This relationship is directly attributed to the kinematics of ball-end milling, implying a variation in the value of the $v_c$ along the active cutting edge. During ball-end milling, the cutting speed value is dependent on the effective diameter of a tool $D_{ef}$, the value of which is smallest at the $P$ point of a cutting edge, while it is highest at the point $M$ (Fig. 11).

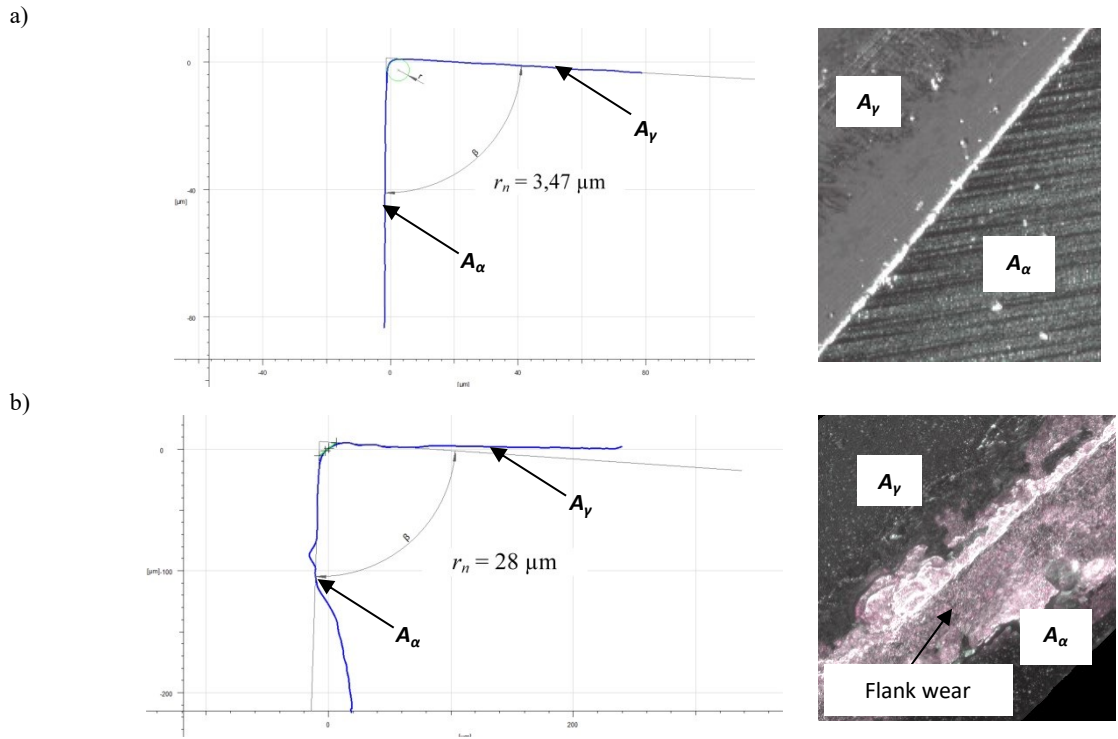

**Fig. 9.** Cutting edge radius and microgeometry of the ball-end mill with: a) $VB_B = 0$, b) $VB_B = 0.15$ mm

The value of the $D_{ef}$ parameter depends on the selected values of the axial cutting depth, the surface inclination angle and the diameter of the tool. During machining with angle $\alpha = 0$, a cutting speed at $P$ point is zero (Fig. 11). The material of a workpiece moving along the cutting edge near point $P$ is not transformed into the chip, but is intensively swollen and ploughed. A consequence of this is the growth of the $h_{min}$ value. The relatively high friction on the flank face (in the area close to the $P$ point of a cutting edge), caused by very low cutting speeds, contribute to an increase in the values of cutting forces (Fig. 7a).

During milling with an angle $\alpha = 15°$, the lowest values of the $h_{min}$ and $k$ parameters are observed. An increase in the $\alpha$ angle from $15°$ to $45°$ induces an increase in the values of the parameters $h_{min}$ and $k$. Considering the variation of the cutting speed $v_c$ along the cutting edge for the above range of angles $\alpha$, on can note that the minimum values of the cutting speed (located at point $E$) are in each case much greater than zero and grow with an increase in the value of the surface inclination angle. Thus, an increase in the value of the $\alpha$ angle implies an increase in the cutting speed $v_c$ (along the length of the active cutting edge, excluding point $M$). This induces an intensification of a heat formed in a machining zone, which causes the growth of plasticization degree of a workpiece in a shearing zones, and thus an increase in the $h_{min}$ value.

**Tab. 2.** Regression equations $F_{tn\_av}$, $F_{rn\_av}$ = f(h) during ball-end milling of AISI L6 steel

| $\alpha$ [°] | Form of equation | Coefficient $R^2$ |
|---|---|---|
| 45 | $F_{tn\_av} = 2{,}34\, h_{ze\_av} + 16{,}8$ | 0.99 |
|  | $F_{rn\_av} = 0{,}54\, h_{ze\_av} + 8{,}7$ | 0.98 |
| 30 | $F_{tn\_av} = 2{,}73\, h_{ze\_av} + 19{,}6$ | 0.99 |
|  | $F_{rn\_av} = 0{,}69\, h_{ze\_av} + 11{,}4$ | 0.99 |
| 15 | $F_{tn\_av} = 2{,}4\, h_{ze\_av} + 14$ | 0.95 |
|  | $F_{rn\_av} = 0{,}85\, h_{ze\_av} + 11{,}7$ | 0.99 |
| 0 | $F_{tn\_av} = 9{,}68\, h_{ze\_av} + 53{,}6$ | 0.99 |
|  | $F_{rn\_av} = 1{,}98\, h_{ze\_av} + 37$ | 0.99 |

It was also observed that in case of ball-end milling of AISI L6 steel, a values of the $k$ parameter (normalized minimum uncut chip thickness) are contained in the range from 0.66 to 0.8. Thus, they slightly exceed the value of $k = 0.63$, obtained during orthogonal turning of hardened alloy steel 1.3505 [35]. These discrepancies can be related to the different grades of hardened steels used in both studies, as well as to a discussed phenomenon of intensive swelling and ploughing of the workpiece material in ball-end milling process with an angle of $\alpha = 0$. An arched profile of the ball-end mill, defining the shape of the instantaneous area of cut, may also be a reason for the differences. As a result, regardless of the considered value of the instantaneous working angle $\varphi$, the area of cut during ball-end milling can be simultaneously divided into the shearing area and the ploughing area (Fig. 12). In this way, the material cut with a curved cutting edge can undergo greater deformation than during machining with a straight cutting edge, which can contribute to an increase in the value of $h_{min}$.

a)
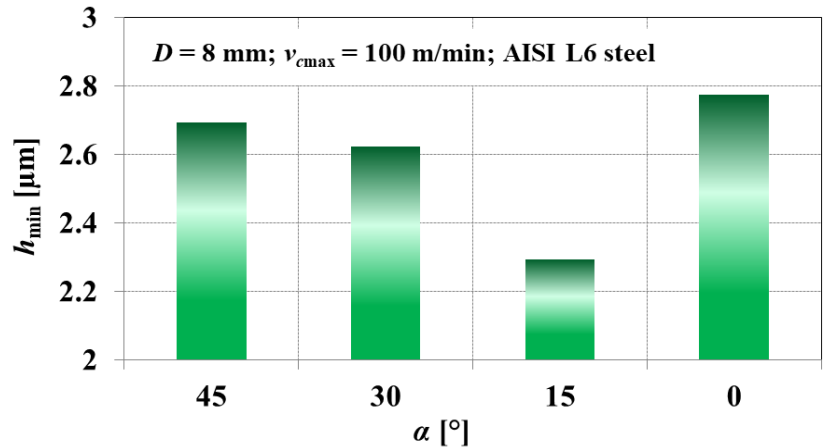

b)
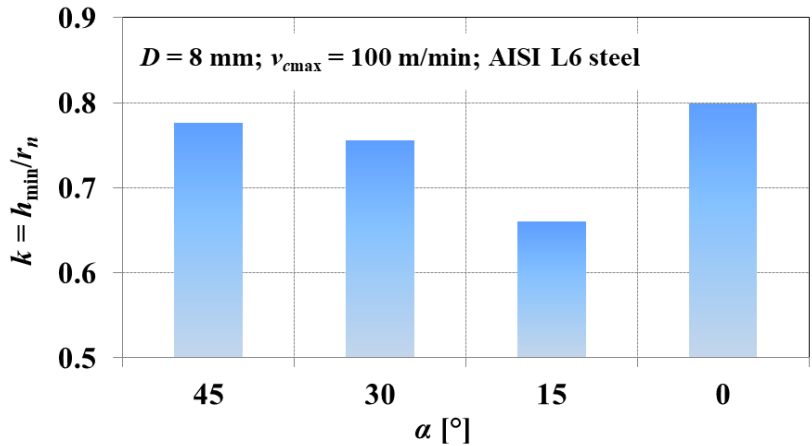

**Fig. 10.** Parameters characterizing material decohesion in ball-end milling of AISI L6 steel: a) $h_{min}$, b) $k$

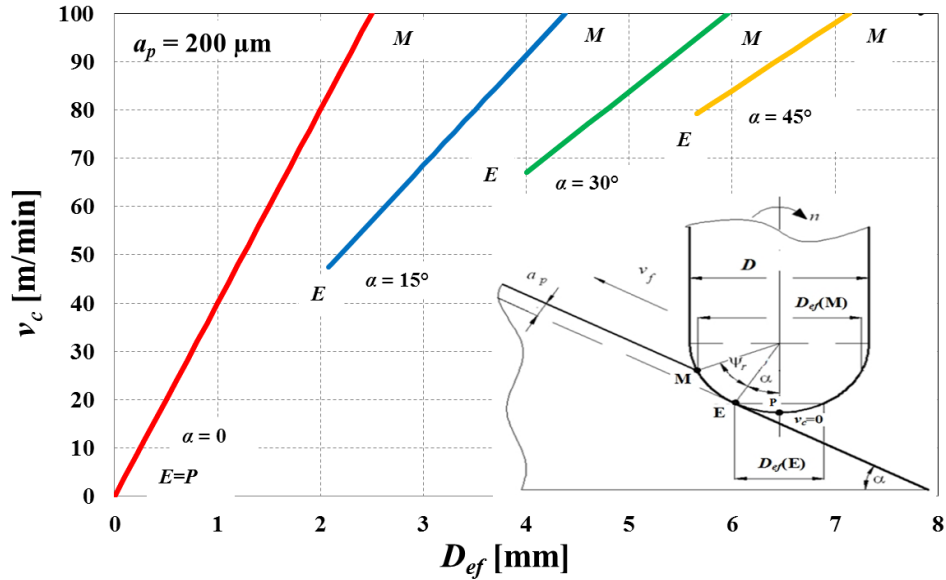

**Fig. 11.** Variation of cutting speed along the active cutting edge of a ball-end mill under machining conditions with $v_{cmax} = 100$ m/min

In order to confirm the hypothesis of the effect of the curvature of a cutting edge on the value of the $h_{min}$ parameter, a theoretical analysis of the developed model was carried out. Its basis was the estimation of $h_{min}$ value for different diameters of the ball-end mill, considering the same range of measured forces. Figure 13 shows that growth of tool nominal diameter $D$ induces a slight decrease in the value of $h_{min}$, which provides analytical confirmation of the relationship between the $h_{min}$ parameter and tool curvature.

Considering a worn cutting tool ($VB_B = 0.15$ mm), it was observed that during machining with an angle $α = 45°$, the value of the $k$ parameter is 0.71. This represents about a nine-percent difference from the value obtained for a new tool ($k = 0.78$). The reason for the differences is the variation in the microgeometry of a cutting edge, affecting the course of thermomechanical and tribological phenomena in the cutting zone. However, despite a slight decrease in the value of the $k$ factor during machining with a worn tool, the $h_{min}$ increased to 19.9 µm. This correlates directly with the intense increase in radius $r_n$ together with a growth of a tool wear.

In order to validate a model of the $h_{min}$ with respect to the ball-end milling, an indirect experimental approach was employed. The method was based on measuring the machined surface roughness in the direction of the feed motion vector $v_f$, and then relating the measured maximum height of surface roughness profile $Rz$ to the theoretical value $Rzt$, determined applying a surface roughness model for a ball-end milling [36]. The model considered the curved mapping of the tool into a workpiece, the run-out of a tool and value of the $h_{min}$ parameter. It was assumed that when $Rz = Rzt$, then the estimated value of $h_{min}$ corresponds to the real value affecting the height of surface roughness (and thus the value of $Rz$). Measurements of surface topography were made on a Veeco NT 1100 optical profile meter, using a 5.1-fold magnification of the analyzed area.

A discrepancies between the measured and experimentally determined values of the $Rz$ parameter are about 4% for machining with an angle of $α = 0°$ and about 18% when milling with $α = 15°$ (Fig. 14). These results demonstrate good accuracy in assessing the $h_{min}$ value with a ball-end mill using the proposed analytic-experimental method. It is worth to mention that the presented model verification method has some limitations due to the complexity of the mechanisms related to the formation of surface roughness during milling. The adapted theoretical model of surface roughness assumes that the formation of the surface profile is only the result of a kinematic-geometric mapping of the ball-end mill into a workpiece, as well as an effect of the $h_{min}$. In practice, the machined surface topography also results from an inhomogeneity of a machined material, a micro-geometry of the tool (related to the cutting edge's honing and progressive wear), and the instant displacements of the tool in relation to the machined material. Differences between measured and calculated values of *Rz* surface roughness may be the result of the interaction of the above-mentioned phenomena, rather than errors in the estimation of $h_{min}$.

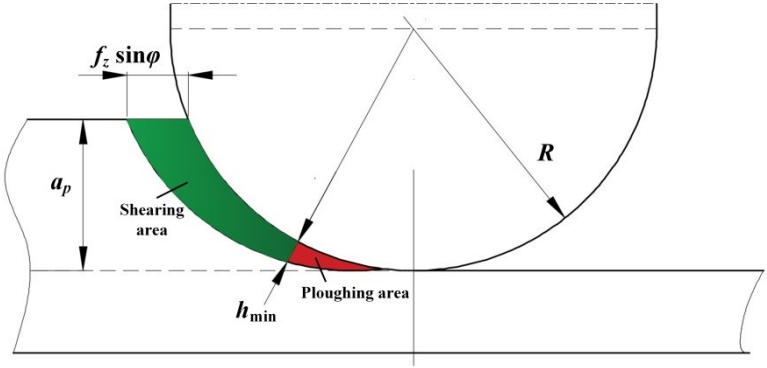

**Fig. 12.** Area of cut in ball-end milling considering the shearing and ploughing areas

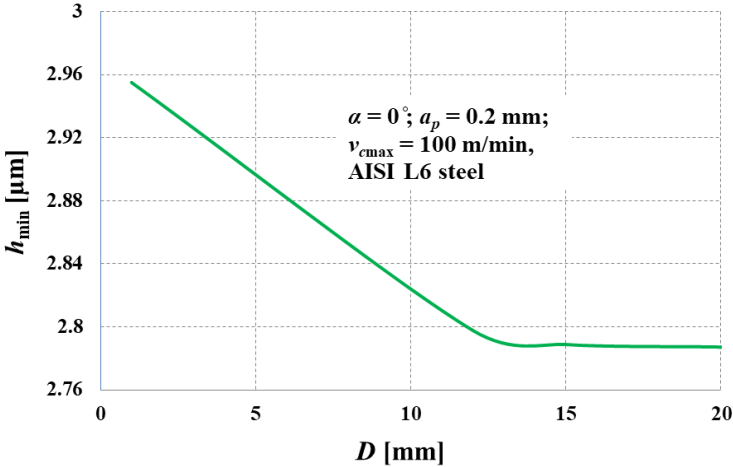

**Fig. 13.** Effect of ball-end mill diameter on the estimated minimum uncut chip thickness

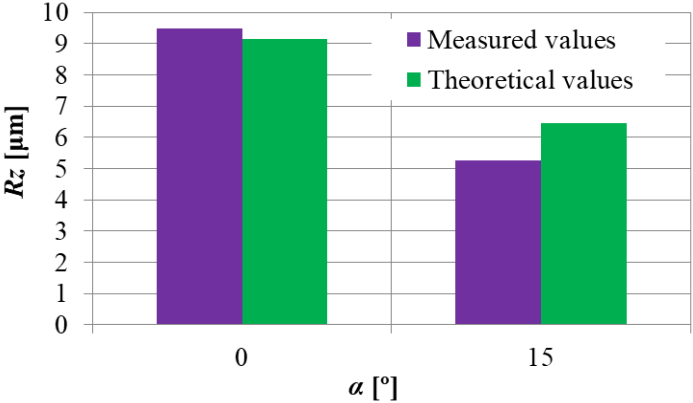

**Fig. 14.** Comparison of measured and theoretical (determined from Brammertz model) surface roughness of hardened AISI L6 steel during ball-end milling

In the next step, the machined 3D surface topographies were evaluated for a slot milling ($\alpha = 0°$) and upward ramping ($\alpha = 15°$). It should be noted that analysis of surface topography is significant in terms of functionality, since it characterizes important properties, as wear resistance, lubricant retention, friction and sealing [37]. The images of three-dimensional surface topography, for machining with angle $\alpha = 0°$ reveal the appearance of arched machining marks, characteristic for milling, as well as clear traces of lateral plastic flow of a workpiece (Fig. 15). The presence of plastic flow is the result of intensive ploughing of a workpiece, which occurs in the range of very low values of cutting speed $v_c$ and results from relatively high value of $h_{min}$. In the case of milling with an angle $\alpha = 15°$, the machined surface lacks the visible machining marks and plastic protrusions (Fig. 16). However, surface irregularities with transverse directionality with respect to the $v_f$ vector were observed parallel to each other. The reason for their occurrence is probably the vibration of the working part of the tool in the radial direction, with frequencies higher than the tooth passing frequency. It is worth noting that these irregularities may have superimposed themselves on the generated machining marks and thus contributed to some differences between the measured values of the $Rz$ parameter and $Rzt$ values estimated from the applied theoretical model.

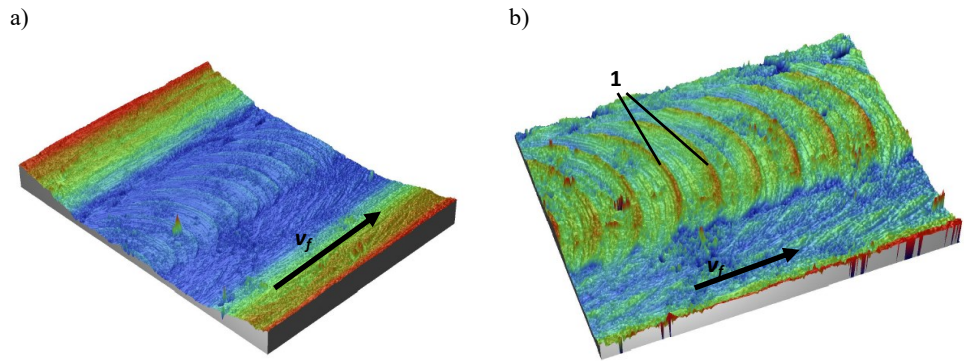

**Fig. 15.** 3D surface topography of the hardened AISI L6 steel after ball-end milling with an angle $\alpha = 0°$: a) analyzed area 0.92 mm x 1.2 mm, b) analyzed area 0.9 mm x 0.89 mm (1 - traces of plastic fash)

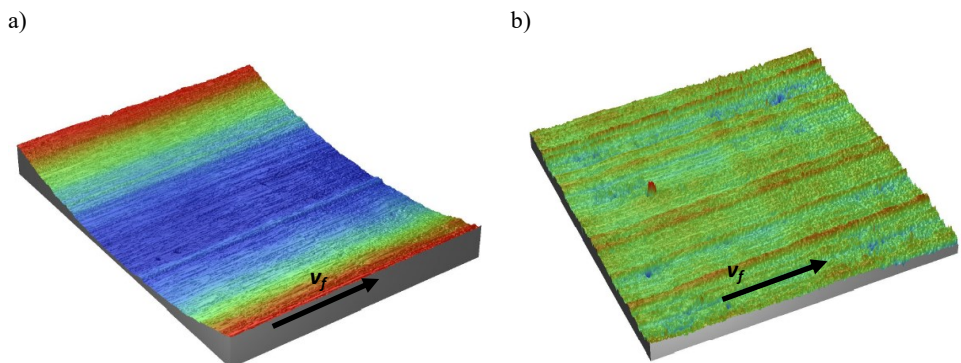

**Fig. 16.** 3D surface topography of the hardened AISI L6 steel after ball-end milling with an angle $\alpha = 15°$: a) analyzed area 0.92 mm x 1.2 mm, b) analyzed area 0.9 mm x 0.89 mm

### 4.2. Evaluation of edge and ploughing forces

The analytic-experimental model formulated as part of research makes it possible to determine the forces acting on a rounded cutting edge of the tool, as well as the cutting force time courses. The developed models were based on a values of proportionality coefficients derived from the expressions shown in table 3. The proportionality coefficients regression models were expressed as a function of surface inclination angle $\alpha$ in a form of a quadratic equations, which stays in accordance with the proportionality coefficients model employed in [33].

Figures 17 and 18 show the courses of forces in the tool coordinate system in the cutting edge radius zone. The values of forces corresponding to $h_z = 0$ and $b > 0$ are the edge forces $F_{te}$, $F_{re}$, $F_{ae}$ associated with frictional phenomena in the flank face-workpiece contact zone. In the range of $0 < h_z \leq h_{min}$, an interval of coupled occurrence of edge forces $F_{te}$, $F_{re}$, $F_{ae}$ and ploughing forces $F_{tp}$, $F_{rp}$, $F_{ap}$ (resulting from the deformation of a workpiece material in relation to the cutting tool below the stagnation point) is observed.

A characteristic feature of the ploughing interval is the nonlinearity of the force in function of uncut chip thickness, observed independently of a milling input parameters. This nonlinearity is due to the inclusion in the force model, the ploughing volume. The results obtained confirm the relationship observed by Bissacco et al. [38]. According to this study, during micromilling of hardened martensitic steel, the measured force values are characterized by a nonlinear course in the range of $h_z/r_n < 0.5$.

Tab. 3. Specific cutting force coefficients equations during ball-end milling of hardened steel AISI L6

| Type of model | Equation |
|---|---|
| Shearing forces | $K_{tc} = 5.42\alpha^2 - 478.9\alpha + 13154$ |
|  | $K_{rc} = 1.13\alpha^2 - 102.7\alpha + 2941$ |
|  | $K_{rc} = 1.4\alpha^2 + 122\alpha - 2131$ |
| Edge forces | $K_{te} = 0.014\alpha^2 - 1.23\alpha + 36.1$ |
|  | $K_{re} = 0.01\alpha^2 - 0.94\alpha + 25$ |
|  | $K_{ae} = -0.01\alpha^2 + 0.65\alpha - 12$ |
| Ploughing forces | $K_{tp} = 1\cdot10^{-4}\alpha^2 - 0.009\alpha + 0.198$ |
|  | $K_{rp} = 2\cdot10^{-5}\alpha^2 - 0.002\alpha + 0.044$ |
|  | $K_{ap} = -4\cdot10^{-5}\alpha^2 + 0.003\alpha - 0.033$ |

Comparing the values of edge forces for different values of surface inclination angles, significant differences are noted. In this case, the obtained values of the edge components during milling with angle $\alpha = 0$ are greater than those obtained in case of milling with $\alpha > 0$. This indicates a significant correlation of frictional phenomena in tool flank face – workpiece interface with the distribution of cutting speeds during machining with different surface inclination angles. The applied surface inclination angles also largely determine the values of the forces occurring in the interval $0 < h_z \leq h_{min}$. The essential factor influencing the ploughing area is the minimum uncut chip thickness. An increase in this parameter contributes directly to an increase in the upper limit of material ploughing, corresponding to the occurrence of the maximum ploughing force.

A comparison of force courses for a ball-end milling with different $\alpha$ angles shows that in the case of machining with $\alpha = 0$, the maximum values of the forces occurring for the $h_z = h_{min}$ are 360% greater than the values obtained during machining with $\alpha = 45°$. This observation confirms the close relationship between a variations of cutting speeds along the active cutting edge and material ploughing phenomenon.

An important factor determining the variation of forces located on a rounded cutting edge of the cutter is the tool flank wear (Fig. 18). During milling with a worn tool ($VB_B = 0.15$ mm), the values of edge forces increased by up to 355% compared to the forces generated during milling with a new tool ($VB_B \approx 0$). This may be related to the increase in contact length between flank face of the tool and workpiece, as well as the loss of material mass resulting from the presence of abrasive and adhesive wear mechanisms. This material loss can affect the formation of grooves/craters along the flank face. An increased contact length between flank face of the tool and workpiece together with the appearance of attrition, as well as grooves/craters on the flank face can lead to the intensified friction in a flank face – workpiece interface and therefore to the increase of edge forces. Considering the forces occurring for uncut chip thickness $h_z = h_{min}$, it is observed that during milling with a worn tool, the maximum values of the $F_r$ force component increased more than 32 times compared to values obtained for a ball-end mill with $VB_B \approx 0$.

The analysis presented above refers to the coupled effect of the edge and ploughing forces appearing in the precision ball-end milling conducted with: $0 \leq h_z \leq h_{min}$. The edge forces are found in the tool flank face – workpiece interface, in the area of $h_z = 0$ and $b > 0$, thus they include mainly the effect of rubbing and friction phenomena. On the other side, the ploughing forces are found in the rounded cutting edge – workpiece interface during machining with uncut chip thicknesses: $0 < h_z \leq h_{min}$ and thus they are mainly responsible for an intense plastic deformations of a workpiece.

Therefore, to characterize the effect of ploughing forces separately from the edge forces, the figures 19 and 20 have been developed. The maximal ploughing forces depicted in figures 19 and 20 were calculated as the absolute values of forces corresponding to minimum uncut chip thickness value ($F_t(h_z = h_{min})$, $F_r(h_z = h_{min})$, $F_a(h_z = h_{min})$) minus the corresponding edge forces ($F_{te}$, $F_{re}$, $F_{ae}$).

a)
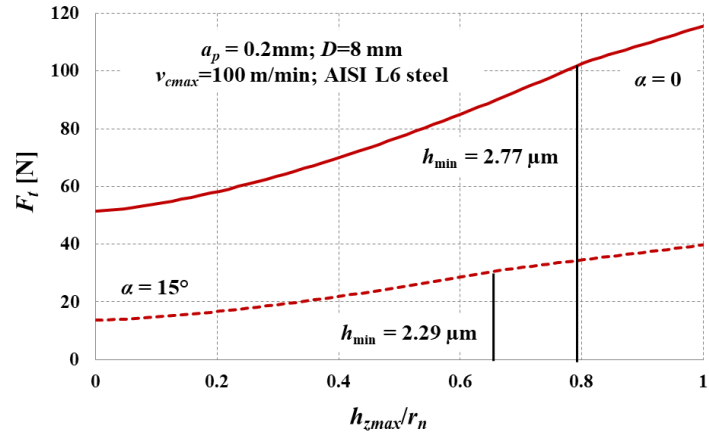

b)
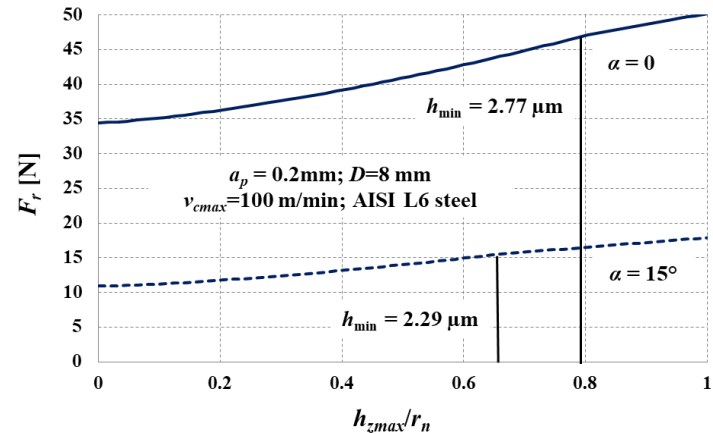

c)
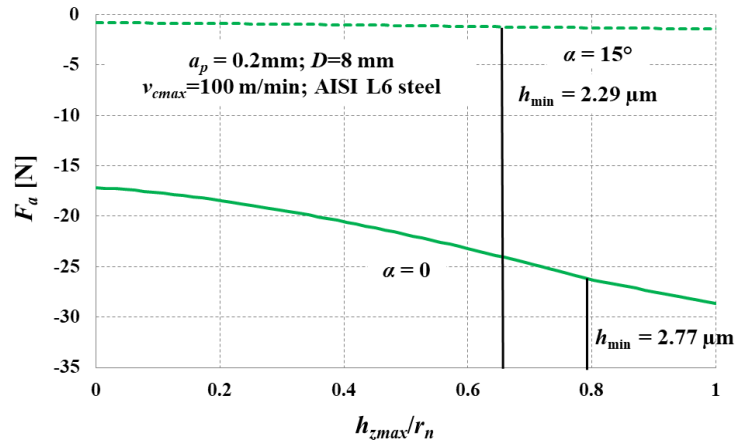

**Fig. 17.** Effect of $h_{zmax}/r_n$ ratio on forces in ball-end milling of hardened AISI L6 steel with $\alpha = 0$ and $\alpha = 15°$: a) tangential force, b) radial force, c) binormal force

From a figure 19 it can be observed that ploughing forces in all investigated directions ($F_{tp}$, $F_{rp}$, $F_{ap}$) have a decreasing trend with an increase in $\alpha$ angle. Thus, the course variations of ploughing forces in a function of machined surface inclination ($\alpha$ angle) are similar to those reached for a $h_{min}$ (see – Fig. 10). It is mainly caused by the directly proportional influence of a maximal ploughing force to the $h_{min}$ value. As it has been stated previously, during ball-end milling with $\alpha = 0$, the cutting speeds in a vicinity of a tool tip are close to zero. Cutting of a hardened steels with a very low cutting speeds can cause the unstable thermomechanical phenomena in the tool-workpiece interface and thus ineffective thermal softening of a workpiece [39]. However, with a growth of a machined surface inclination (cutting speed in the $E$ point of a cutting edge – see Fig. 11), the thermomechanical relations between a tool and workpiece are stable. This relation can contribute to the decrease in ploughing forces.

a)
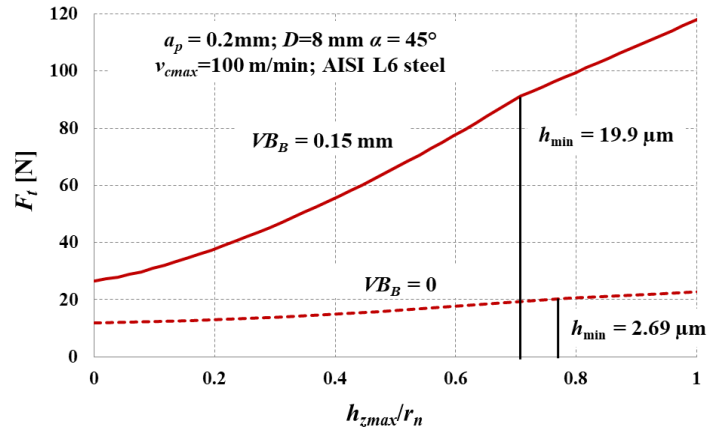

b)
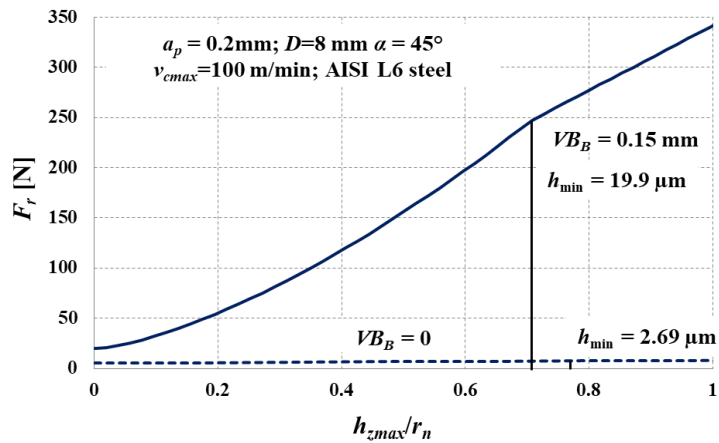

c)
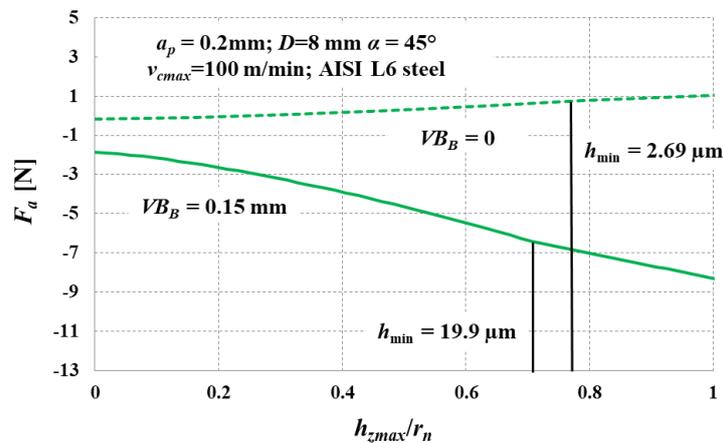

**Fig. 18.** Effect of $h_{zmax}/r_n$ ratio on forces in ball-end milling of AISI L6 steel with $VB_B = 0$ and $VB_B = 0.15$mm: a) tangential force, b) radial force, c) binormal force

It can be also observed that independently on the tested $\alpha$ angle, the highest ploughing forces are reached in tangential direction $F_{tp}$, than in the radial one $F_{rp}$ and finally the lowest values are reached for a binormal direction $F_{ap}$. The highest values of ploughing forces in the tangential direction can be explained by a collinearity of the $F_{tp}$ force vector with a vector of a cutting speed, which consequently leads to a most intense ploughing action. On the other hand, the binormal direction of force is strictly correlated with a ball-end mill helical geometry (affected by the helix angle). Theoretically, in ball-end milling with $\lambda_s = 0$ no forces in binormal direction should appear. However, in case of ball-end milling with a non-zero helix angle $\lambda_s \neq 0$, the forces affected by a lateral plastic deformation of a workpiece can be found and their values are thus lower than those in tangential and radial directions (which in turn are influenced by a flow of a workpiece against the tool with a linear speed equal to the cutting speed value).

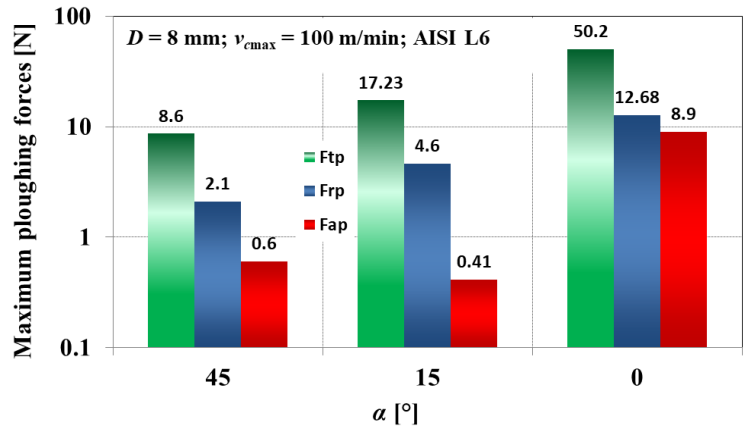

**Fig. 19.** The ploughing forces appearing in ball-end milling with various surface inclination angles

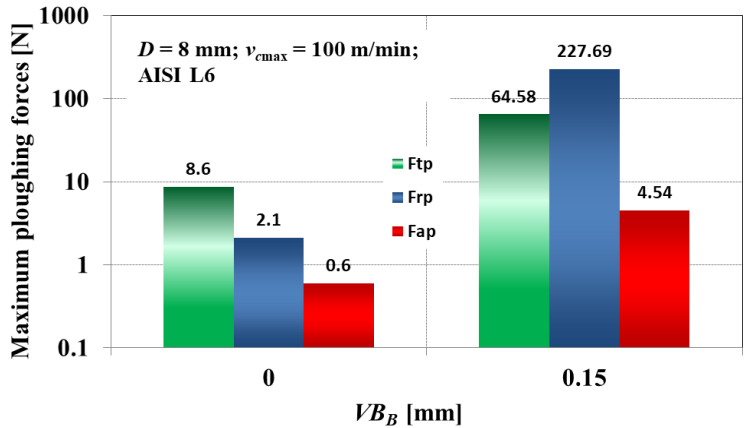

**Fig. 20.** The ploughing forces appearing in ball-end milling with various tool wear and $\alpha = 45°$

Figure 20 shows the intense influence of a tool wear on generated values of ploughing forces in all investigated directions ($F_{tp}$, $F_{rp}$, $F_{ap}$). During milling with a worn tool ($VB_B = 0.15$ mm) the ploughing force components increased 7.5-fold (in tangential and binormal directions), and even 108-fold in case of radial direction, comparing to corresponding values in ball-end milling with a new tool. These vast differences prove that progressing tool wear has a tremendous influence on the ploughing phenomena during the chip decohesion. Thus, tool wear has a major effect on the course of a ploughing phenomenon, contributing significantly to the increase in the ploughing force components $F_{tp}$, $F_{rp}$, $F_{ap}$. This is caused to the greatest extent by a change in the microgeometry of a cutting edge profile, corresponding directly to a progressing tool wear (e.g. a significant growth of cutting edge radius $r_n$ – Fig. 9, or alteration in shape of cutting edge leading to the growth of friction at the cutting edge radius-workpiece interface). These changes can induce an increase of the uncut chip thickness in the ploughing regime, thereby increasing the ploughing volume. The highest values of radial ploughing components $F_{rp}$ reached during milling with a worn tool can be attributed to the irregular and non-circular cutting edge profile, as well as the generation of microgrooves on the cutting edge in the area below the stagnant point, as well as in a flank face (Fig. 21). This can consequently lead to an intense friction and thermomechanical interactions in tool-workpiece interface. In order to characterize the relationships between a progressing tool wear and ploughing forces, a SEM images of ball-end mills were presented (Fig. 22).

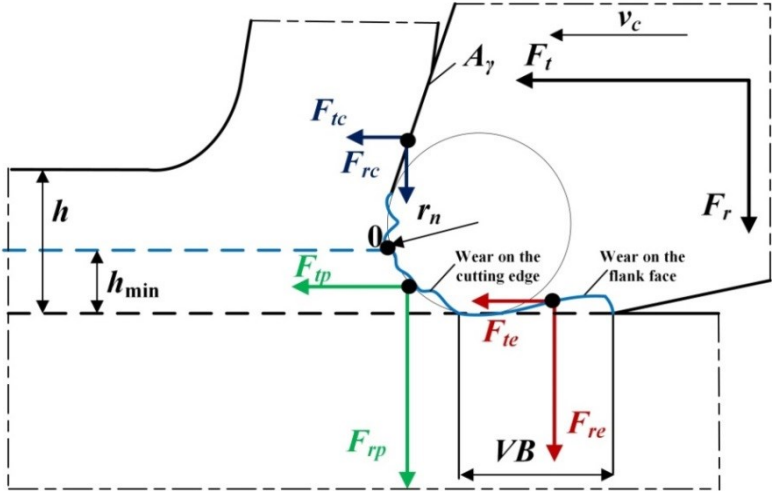

**Fig. 21.** Model depicting distribution of shearing, edge and ploughing forces for a worn tool

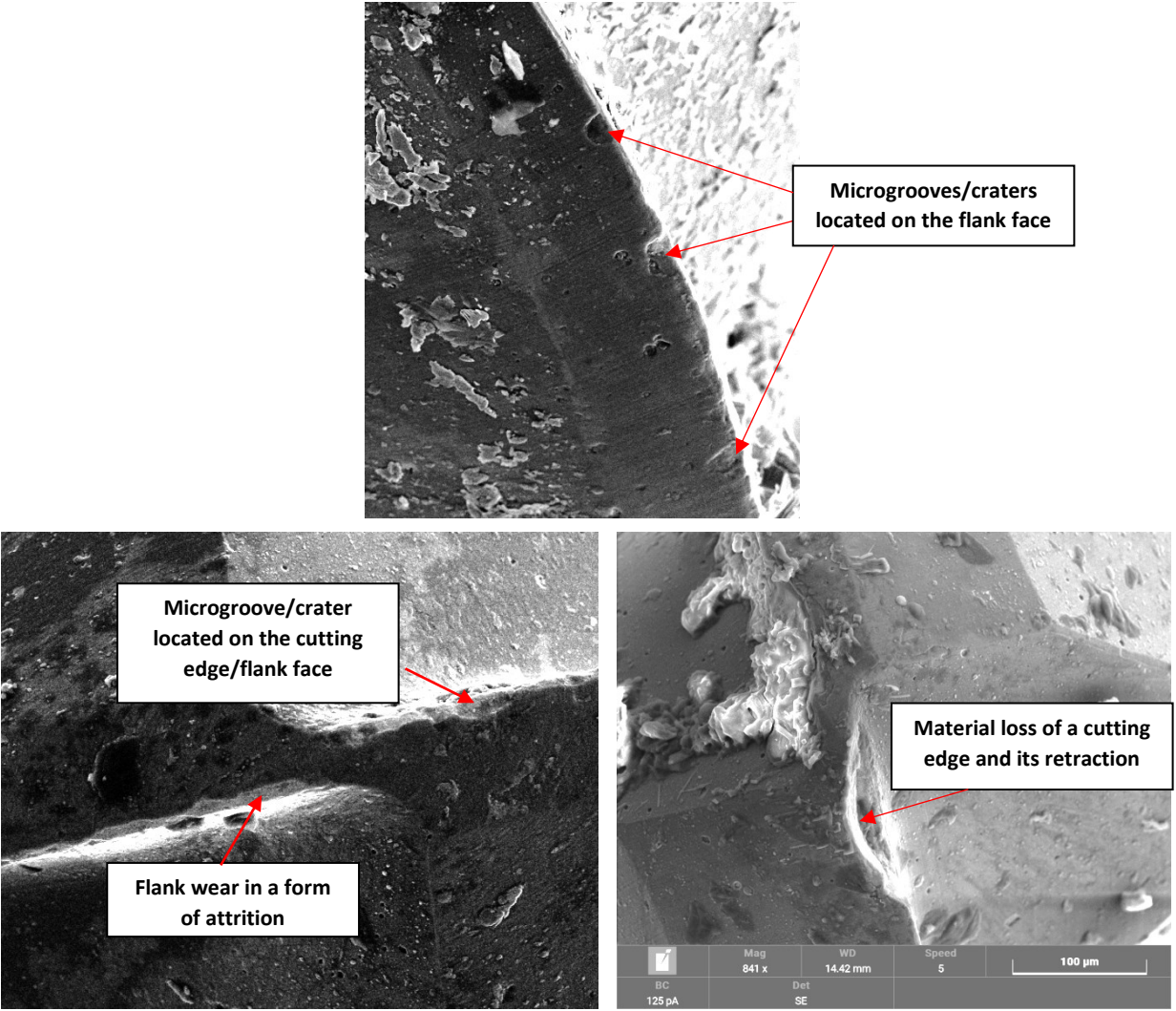

**Fig. 22.** Tool wear of the ball-end mills during milling with $\alpha = 45°$

The SEM images of a ball-end mills for the preliminary tool wear ($VB_B < 0.05$ mm) show that during milling process of a hardened steel, the abrasive tool flank wear together with a local wear in a form of a microgrooves are found. The abrasive tool wear is mainly caused by a friction in tool-workpiece interface. However, the appearance of microgrooves can be induced by a local exceeding of ultimate strength of tool material, which in turn can be caused by some cutting force oscillations and adhesion wear mechanism. This in turn can affect the edge and ploughing forces. It should be stated here that abrasion and adhesion are the most common wear mechanisms during precise milling with solid carbide mills of hard-to-cut materials. Sen et al. [40] indicated that adhesion and abrasion are the typical wear mechanisms of TiAlN coated solid carbide end mills during milling of Inconel 690. Similar findings were obtained by Sousa et al. [41] in the study focused on evaluating the tool wear during milling of W 1.2711 pre-hardened tool steel with TiAlSiN and TiAlN PVD coated solid carbide end mills.

The SEM images show that progressing tool wear can induce the material loss of a rounded cutting edge and changes in its shape. These observations confirm the alterations of ball-end mill microgeometry, induced by a progressing tool wear, which are manifested in a form of material loss, abrasion, as well as microgrooves.

The next step in the analysis of the formulated models focused on the time courses of forces. Figure 23 shows the time courses of ball-end milling forces for $h_{min}/f_z = 0.034$, estimated from the developed force models and determined using the measured values of cutting force components (according to the relation: $F = (F_x^2 + F_y^2 + F_z^2)^{0.5}$). In order to estimate the forces, two different approaches were used that differed in the way the proportionality coefficients were determined. In the first approach, the influence of ploughing forces was not considered (the effect of ploughing coefficients was omitted), and only shearing and edge coefficients were determined. It should be noted that this method is most commonly used to determine the value of proportionality coefficients during ball-end milling of curved surfaces [42]. In the second approach, the effect of ploughing forces was also taken into account. Due to the non-monotonic effect of $\alpha$ angle on the values of proportionality coefficients, regression models were used in the form of a quadratic equation (Tab. 3).

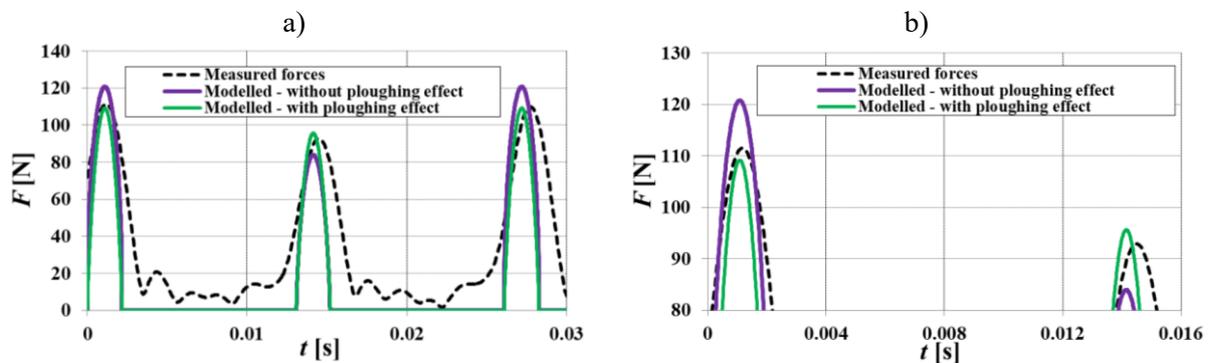

**Fig. 23.** The time courses of a resultant cutting force in ball-end milling of AISI L6 steel with $h_{min}/f_z = 0.034$ and $\alpha = 45°$: a) in a range of a one full cutting tool rotation, b) magnified view depicting milling with 2 teeth

From the analysis of the cutting force time courses presented in figure 23, it is clear that the maximum forces per tooth take alternately higher and lower values. This phenomenon can be attributed to the radial runout of the tool (in the analyzed case amounting to 3 µm) determining the actual values of the area of the cut (equation (20)). During precision machining with $\alpha > 0$, pulsating force time courses are observed. For 2-teethed cutters, this phenomenon occurs when the working angle of the cutter is less than 180°.

The measured cutting force signal (see – Fig. 23a) reveals also an appearance of some fluctuations with a frequencies higher than a tooth passing frequency. These fluctuations can be induced by the effective uncut chip thickness variations resulting from vibrations occurring in the machining system, discontinuities of a workpiece structure, pre-machining surface irregularities, uncut chip thickness accumulation and other factors. A variations of an effective uncut chip thickness below the minimum uncut chip thickness value ($h < h_{min}$) can cause the alterations of a ploughing volume, and thus the ploughing forces. A variable ploughing forces (during one tool immersion into the workpiece) can lead to a variations of dynamic damping coefficient between the workpiece and tool flank face (contributing to a dynamic stability of milling process), as well as to the growth of a tool wear. However, in order to characterize the instant variations of a ploughing forces during milling, the dynamic ploughing force model (including – among others – the uncut chip thickness regeneration mechanism) should be further developed.

The resultant force time courses, determined on the basis of the model taking into account the effect of ploughing forces show greater agreement with experimental force signals compared to those estimated using only the shear and edge coefficients. The error in the estimation of maximum forces, per tooth, was about 5% for the model including ploughing forces and 18.7% for the model without the influence of ploughing forces. Thus, it can be concluded, that the effect of ploughing forces has an effect on the modelled forces not only during the micro ball-end milling (e.g. when $f_z < r_n$) but also during precise cutting with $f_z \approx r_n$ and $f_z > r_n$.

## 5. CONCLUSIONS

Research conducted in this study involved the formulation of $h_{min}$ model for a ball-end milling of AISI L6 alloy hardened steel, based on the identification of a stagnant point. Additionally, a ploughing force model was proposed, applying the identification of a ploughing constants and ploughing volume obtained from analytical expressions and cutting force measurements. Based on conducted calculations and measurements, the following conclusions were formulated.

- The $\alpha$ angle, characterizing the machined surface inclination has a non-monotonic influence on the estimated values of $h_{min}$ and $k$. The highest values of $h_{min}$ and $k$ were reached during the slot ball-end milling with $\alpha = 0$. Contrarily, a lowest $h_{min}$ and $k$ values were obtained during milling with $\alpha = 15°$. Further increase in surface inclination angle led to the increase of $h_{min}$ and $k$.
- During the milling with a worn tool ($VB_B = 0.15$ mm), the $k$ parameter decreased by about 9% comparing to the value reached during milling with a new tool. However, apart from the decrease of a $k$ parameter, the $h_{min}$ value increased, which is directly correlated with an intense growth of a cutting edge radius during ball-end milling with a worn tool.
- Validation of the $h_{min}$ estimation method applying a surface roughness model (considering the kinematic – geometric projection and $h_{min}$ value) reveals that the differences between the measured and experimentally determined values of the $Rz$ parameter are approximately 4% for machining with a surface inclination angle $\alpha = 0$ ° and approximately 18% for milling with $\alpha = 15°$. These results prove the good accuracy of estimating the $h_{min}$ during ball-end milling using the proposed analytic-experimental method.
- Ploughing phenomenon during milling manifests as the the non-linearity of forces in function of the uncut chip thickness, and it is observed regardless of the value of the surface inclination angle and the condition of a milling tool. This nonlinearity results from taking into account the ploughing volume in a force model. In the case of machining with $\alpha = 0$, the maximum values of the ploughing forces are even 360% higher than the values obtained during machining with $\alpha = 45°$. This observation confirms the close relationship between the distribution of the $v_c$ parameter on the active cutting edge and the material ploughing phenomenon.
- An important factor determining the variations of forces located on the rounded cutting edge of cutter is a tool flank wear. Considering the ploughing forces, it is observed that during milling with a worn tool, the maximum values of radial forces increased 108-fold compared to the values obtained during milling with a new tool. Therefore, the flank wear, and the wear of a cutting edge have a major impact on the ploughing phenomenon intensity, contributing significantly to the increase of force components.

- Resultant force time courses estimated on the basis of a model taking into account the influence of the ploughing forces show greater compliance with the experiment compared to the values simulated using only the shear and edge coefficients. The error in estimating the maximum forces was about 5% for the model including ploughing forces and 18.7% for the model without the effect of ploughing phenomenon. This observation substantiates the application of the accurate force models with a consideration of ploughing phenomenon during precise ball-end milling with the $h_z > r_n$.

**Acknowledgments**

This work was supported by the National Centre of Science (Decision No. 2017/25/B/ST8/00962) "Modeling of dynamics and strength problems during precision milling with micro ball end mills" and National Centre of Science (Decision No. 2021/43/D/ST8/00023) "Modeling and research on surface texture formation during ultra-precision micromilling".